\documentclass{article}
\usepackage{graphicx}
\usepackage{comment}
\usepackage{amsmath,amssymb}
\usepackage{color}
\usepackage{url}
\usepackage{marvosym}
\usepackage{hyperref}
\hypersetup{
    colorlinks,
    linkcolor={red!50!black},
    citecolor={blue!50!black},
    urlcolor={blue!80!black}
}
\usepackage[dvipsnames, x11names]{xcolor}
\usepackage{booktabs}
\usepackage[left=3cm,right=3cm,top=3cm,bottom=3cm]{geometry}
\usepackage{multirow}
\usepackage{glossaries-extra}
\glsdisablehyper
\usepackage{xr}
\usepackage{todonotes}
\newabbreviation{mri}{MRI}{magnetic resonance imaging}
\newabbreviation{nmr}{NMR}{nuclear magnetic resonance}
\newabbreviation{sde}{SDE}{stochastic differential equation}
\newabbreviation{pc}{PC}{predictor--corrector}
\newabbreviation{tv}{TV}{total variation}
\newabbreviation{vp}{VP}{variance preserving}
\newabbreviation{ve}{VE}{variance exploding}
\newabbreviation{mcmc}{MCMC}{Markov chain Monte Carlo}
\newabbreviation{hmc}{HMC}{Hamiltonian Monte Carlo}
\newabbreviation{corpd}{CORPD}{coronal proton-density weighted}
\newabbreviation{corpdfs}{CORPDFS}{coronal proton-density weighted and fat suppressed}
\newabbreviation{psnr}{PSNR}{peak signal-to-noise ratio}
\newabbreviation{ssim}{SSIM}{structural similarity index measure}
\newabbreviation{qm}{QM}{quantum mechanics}
\newabbreviation{cm}{CM}{classical mechanics}
\newabbreviation{gan}{GAN}{generative adversarial network}
\newabbreviation{gfp}{GFP}{Gaussian Fourier projection}
\newabbreviation{erf}{ERF}{effective receptive field}
\newabbreviation{p}{P}{Poisson disk sampled mask}
\newabbreviation{r}{R}{Radial mask}
\newabbreviation{g}{G-1D}{1D Gaussian mask}
\newabbreviation{g2}{G-2D}{2D Gaussian mask}
\newabbreviation{rf}{RF}{radio frequency}
\newabbreviation{ood}{OOD}{out-of-distribution}
\newabbreviation{id}{ID}{in-distribution}
\newabbreviation{pi}{PI}{parallel imaging}
\newabbreviation{cs}{CS}{compressed sensing}
\newabbreviation{squid}{SQUID}{superconducting quantum interference device}
\newabbreviation{gpu}{GPU}{graphics processing unit}
\newabbreviation{fid}{FID}{free induction decay}
\newabbreviation{fov}{FOV}{field of view}
\newabbreviation{vn}{VN}{variational network}
\newabbreviation{modl}{MoDL}{model-based deep learning architecture}
\newabbreviation{foe}{FoE}{field of experts}
\newabbreviation{em}{EM}{Euler--Maruyama}
\newabbreviation{iid}{i.i.d.}{independent and identically distributed}
\newabbreviation{ncsn}{NCSN}{noise conditional score network}
\newabbreviation{ald}{ALD}{annealed Langevin dynamics}
\newabbreviation{silu}{SiLU}{sigmoid-weighted linear unit}
\newabbreviation{zf}{ZF}{zero-filled}
\newabbreviation{gn}{GN}{group norm}
\newabbreviation{af}{AF}{acceleration factor}
\newabbreviation{ct}{CT}{computed tomography}

\usepackage{amssymb}
\usepackage{mathtools}
\usepackage[ruled,noend,linesnumbered]{algorithm2e}
\usepackage{siunitx}
\usepackage{microtype}
\usepackage{ifthen}
\usepackage{tikz}
\usepackage{pgfplots}
\pgfplotsset{compat=1.18} 
\usetikzlibrary{positioning, spy}
\SetKw{Continue}{continue}

\DeclareMathOperator*{\argmin}{arg\,min}

\DeclarePairedDelimiter\norm{\lVert}{\rVert}
\usepgfplotslibrary{groupplots}

\allowdisplaybreaks

\pgfplotsset{%
	colormap={inferno}{%
		rgb = (1.46200e-03, 4.66000e-04, 1.38660e-02)
		rgb = (2.94320e-02, 2.15030e-02, 1.14621e-01)
		rgb = (9.29900e-02, 4.55830e-02, 2.34358e-01)
		rgb = (1.83429e-01, 4.03290e-02, 3.54971e-01)
		rgb = (2.71347e-01, 4.09220e-02, 4.11976e-01)
		rgb = (3.60284e-01, 6.92470e-02, 4.31497e-01)
		rgb = (4.41207e-01, 9.93380e-02, 4.31594e-01)
		rgb = (5.28444e-01, 1.30341e-01, 4.18142e-01)
		rgb = (6.09330e-01, 1.59474e-01, 3.93589e-01)
		rgb = (6.94627e-01, 1.95021e-01, 3.54388e-01)
		rgb = (7.69556e-01, 2.36077e-01, 3.07485e-01)
		rgb = (8.41969e-01, 2.92933e-01, 2.48564e-01)
		rgb = (8.98192e-01, 3.58911e-01, 1.88860e-01)
		rgb = (9.44285e-01, 4.42772e-01, 1.20354e-01)
		rgb = (9.72590e-01, 5.29798e-01, 5.33240e-02)
		rgb = (9.86964e-01, 6.30485e-01, 3.09080e-02)
		rgb = (9.84865e-01, 7.28427e-01, 1.20785e-01)
		rgb = (9.66243e-01, 8.36191e-01, 2.61534e-01)
		rgb = (9.46392e-01, 9.30761e-01, 4.42367e-01)
		rgb = (9.88362e-01, 9.98364e-01, 6.44924e-01)
	},
	colormap={mycolors}{
		color=(MidnightBlue)
		color=(Peach)
		color=(JungleGreen)
		color=(Maroon)
		color=(OliveGreen)
		color=(Plum)
	},
}
\newcommand{\drawcolorbar}{%
	\pgfplotscolorbardrawstandalone[
		scale=0.32, colormap={example}{samples of colormap = (8 of inferno)},
		colorbar horizontal,point meta max=0.2,colorbar style={ticks=none},
	]%
}

\pgfplotscreateplotcyclelist{mycolors}{
	{color=MidnightBlue},
	{color=Peach},
	{color=JungleGreen},
	{color=Maroon}, 
    {color=OliveGreen}, 
    {color=Plum}
}
\pgfplotscreateplotcyclelist{mycolorsmark}{
	{color=MidnightBlue, mark=x},
	{color=Peach, mark=x},
	{color=JungleGreen, mark=x},
	{color=Maroon, mark=x}, 
	{color=OliveGreen, mark=x},
	{color=MidnightBlue, dashed, mark=x, mark options={solid}},
	{color=Peach, dashed, mark=x, mark options={solid}},
	{color=JungleGreen, dashed, mark=x, mark options={solid}},
	{color=Maroon, dashed, mark=x, mark options={solid}}, 
	{color=OliveGreen, dashed, mark=x, mark options={solid}}
}
\pgfplotscreateplotcyclelist{mycolorsfill}{
	{fill=MidnightBlue, draw=MidnightBlue},
	{fill=Peach, draw=Peach},
	{fill=JungleGreen, draw=JungleGreen},
	{fill=Maroon, draw=Maroon}, 
    {fill=OliveGreen, draw=OliveGreen}
}
\DeclareSIUnit{\arbitraryunit}{a.u.}
\usepackage{xspace}
\usepackage[capitalise, noabbrev]{cleveref}
\crefname{equation}{Eq.}{Eqs.}
\newcommand\etal{et al.}
\newcommand\ie{i.e.}
\newcommand\eg{e.g.}

\title{Bigger Isn’t Always Better: Towards a General Prior\\for Medical Image Reconstruction}
\author{%
	Lukas Glaszner\thanks{Institute of Computer Graphics and Vision, Graz University of Technology,\\Inffeldgasse 16/II, 8010 Graz, Austria, \texttt{lukas.glaszner@student.tugraz.at}} \and %
	Martin Zach\thanks{Biomedical Imaging Group, École polytechnique fédérale de Lausanne, 1015 Lausanne, Switzerland and Center for Biomedical Imaging, 1015 Lausanne, Switzerland, \texttt{martin.zach@epfl.ch}}}

\begin{document}
\maketitle
\begin{abstract}
    Diffusion model have been successfully applied to many inverse problems, including MRI and CT reconstruction.
    Researchers typically re-purpose models originally designed for unconditional sampling without modifications.
    Using two different posterior sampling algorithms, we show empirically that such large networks are not necessary.
    Our smallest model, effectively a ResNet, performs almost as good as an attention U-Net on in-distribution reconstruction, while being significantly more robust towards distribution shifts.
    Furthermore, we introduce models trained on natural images and demonstrate that they can be used in both MRI and CT reconstruction, out-performing model trained on medical images in out-of-distribution cases.
    As a result of our findings, we strongly caution against simply re-using very large networks and encourage researchers to adapt the model complexity to the respective task.
    Moreover, we argue that a key step towards a general diffusion-based prior is training on natural images.
\end{abstract}
\section{Introduction}
\label{sec:intro}
Generative deep learning approaches, in particular diffusion models \cite{Chang2023,Song2021DDIM,Ho2020}, have made significant advances over the last several years.
In addition to generating images from scratch, they have also been successfully applied to various inverse problems, such as image in-painting~\cite{Song2021Score,Chung2023,Kawar2022,Lugmayr2022}, super-resolution~\cite{Gao2023,Kawar2021} or medical image reconstruction~\cite{Zach2021,Zach2023,Song2022,Chung2022Score}.

\begin{figure}[th]
    \centering
    \def\wwidth{4}
    \resizebox{\textwidth}{!}{\begin{tikzpicture}
        \foreach [count=\ii] \dataset in {corpd, corpdfs, brain, ct, ct_head, celeba}
        {
            \pgfmathsetmacro{\xpos}{\wwidth*mod(\ii-1, 3)}
            \pgfmathsetmacro{\ypos}{-\wwidth*(floor((\ii - 1) / 3) + 1)}
            \pgfmathsetmacro{\cmapindex}{\ii-1}
            \node [line width=4pt, rectangle, inner sep=0, index of colormap={\cmapindex of mycolors}, color=., draw] at (\xpos cm, \ypos cm) {\includegraphics[width=3.8 cm]{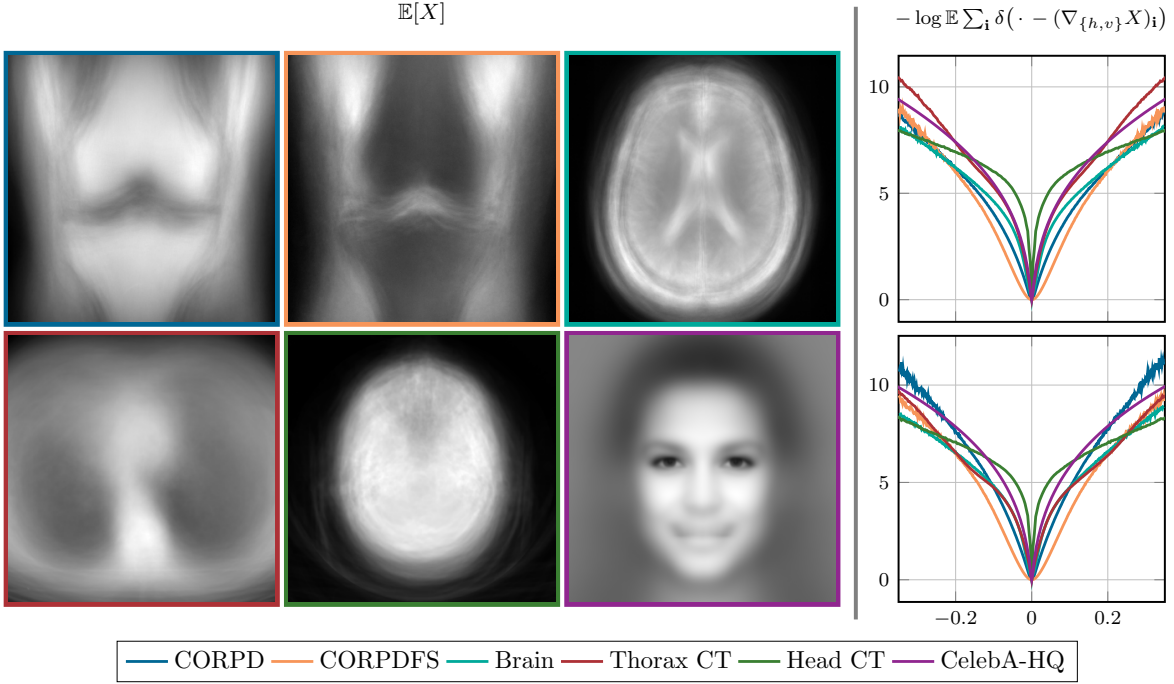}};
        }
        \node at (4cm, -1.5cm) {\small \( \mathbb{E}[X] \)};
			\draw [ultra thick, gray] (10.19, -1.4) -- ++(0, -8.75);
			\begin{scope}[xshift=10.8cm, yshift=-5.9cm]
				\begin{groupplot}[
					group style={
						group size=1 by 2,
						x descriptions at=edge bottom,
						vertical sep=.2cm
					},
					height=3.8cm,
					xmin=-0.35,
					xmax=0.35,
					width=3.8cm,
					scale only axis,
					cycle list name=mycolors,
					no markers,
					grid=major,
					table/col sep=comma,
					ticklabel style={font=\footnotesize},
					every axis plot/.append style={line width=0.4mm},
					]
					\nextgroupplot[thick, title={ \footnotesize \(-\log \mathbb{E}\sum_{\mathbf{i}} \delta\bigl(\,\cdot\, - (\nabla_{\{h, v\}} X)_{\mathbf{i}}\bigr) \)}]
					\foreach \dataset in{0,...,5}
					{
						\addplot table{./data/hist_x_\dataset.csv};
					}
					\nextgroupplot[thick, legend to name = leg, legend columns=-1]
					\foreach \dataset in{0,...,5}
					{
						\addplot table{./data/hist_y_\dataset.csv};
					}
					\addlegendentry{CORPD}
					\addlegendentry{CORPDFS}
					\addlegendentry{Brain}
                    \addlegendentry{Thorax CT}
                    \addlegendentry{Head CT}
					\addlegendentry{CelebA-HQ}
				\end{groupplot}
			\end{scope}
   \node at (6.5, -10.75) {\pgfplotslegendfromname{leg}};
    \end{tikzpicture}}
    \vspace{-8mm}\caption{%
        Left: The means over the six datasets.
        Right: The negative-log histograms of the horizontal (top) and vertical (bottom) image gradients.%
    }%
    \label{fig:mean_images}
\end{figure}

In undersampled \gls{mri} reconstruction, for example, deep learning--based approaches have superseded purely hand-crafted priors such as $\ell_1$-wavelet compressed sensing~\cite{Lustig2007} or \gls{tv}~\cite{Knoll2011} years ago. At first, research has mainly focused on discriminative approaches \cite{Hammernik2018,Aggarwal2018,Schlemper2017}, which are able to generate remarkably good reconstructions. However, a key disadvantage of these models is that they typically fail to generalize across contrasts, anatomies and undersampling masks and have to be trained separately for each combination of these factors. This is an important obstactle to their adoption, because high-quality large data sets are only available for very few anatomies. Generative models overcome this limitation and exhibit an incredible robustness towards distributions shifts~\cite{Zach2023,Luo2023,Chung2022Score,Jalal2021}.
They also offer other advantages, such as the ability to estimate the uncertainty behind a reconstruction---these benefits make generative approaches more viable for clinical adoption in the long run.

Despite the promising results, there are still some open questions. The reconstruction process of these models is typically described as posterior sampling, where the goal is, given the unconditional distribution $p_{X_0}$ and a measurement model $p_{Y|X_0}$, to generate samples from $p_{X_0|Y}$. However, one can easily see that this cannot be the entire story, because in \gls{ood} reconstructions, the network recovers signals from a distribution it has not been trained on at all---this should clearly be impossible. Moreover, even in in-distribution (ID) settings, true posterior sampling is impossible for most models and computationally intractable for all others, as Gupta~et al.~\cite{Gupta2024} have recently shown. 

If they do not perform posterior sampling, then how do these algorithms reconstruct undersampled medical images?
On a global scale, \gls{mri} and \gls{ct} images of different body parts and contrasts differ greatly from each other and from natural images.
Locally however, they follow a similar distribution.
In order to demonstrate this, let $X$ denote a random variable described by one of six data sets: 
proton density weighted knee \gls{mri}, proton density weighted fat suppressed knee \gls{mri}, brain \gls{mri}, thoracic and head \gls{ct} and images of the faces of celebrities.
In \cref{fig:mean_images}, we show the mean $\mathbb{E} [ X ]$ as a global statistic and the negative log-histogram of the image gradient \cite{Huang1999}, \ie 
\begin{equation}
    -\log \mathbb{E}\sum_{\mathbf{i}} \delta\bigl(\,\cdot\, - (\nabla_{\{h, v\}} X)_{\mathbf{i}}\bigr),
\end{equation}
as a local statistic.
Here, $\mathbf{i}$ is a two-index ranging over all pixels and \( \nabla_h \) and \( \nabla_v \) are horizontal and vertical first order finite difference operators, see \eg~\cite{Chambolle2004}.
While the data sets are distinct globally, the local statistics are very similar.

This leads us to hypothesize that at least the out-of-distribution reconstructions are primarily driven by local features, which share similarities across contrasts and structures. If this were the case, it would have several important implications:
First, that the large models which are typically used by default are not actually necessary; this in turn could significantly reduce both the training and reconstruction times.
Second, that we can design more robust \gls{mri} and \gls{ct} reconstruction algorithms by utilizing smaller models.
And third, that models which work on a local scale could be trained on vastly different data sets, such as natural images, as long as local image statistics are similar.

In this work, we set out to test these theories;
to that end, we modify the well-known score-based generative model~\cite{Song2021Score}, which has previously been successfully applied to medical image reconstruction~\cite{Chung2022Come}, such that it has differently-sized receptive fields.
Then, we train these models on \gls{mri}, \gls{ct} and natural images and subsequently evaluate them on different medical image datasets, both in- and out-of-distribution.

\section{Methods}

For the sake of simplicity, throughout this paper we stick to real-valued simulation studies.
Thus, we assume that the imaged anatomy is real-valued and construct the data via retrospective undersampling.
Formally, for an image \( x \in \mathbb{R}^n \) of size \( n = n_1 \times n_2\), the data are given by \( y = Ax \) where the forward operator \( A \) depends on the imaging modality. In case of \gls{mri} \( A : \mathbb{R}^n \to \mathbb{C}^n : x \mapsto M\mathcal{F}x \) combines the two-dimensional discrete Fourier transform \( \mathcal{F} \) and the binary mask \( M = \operatorname{diag}(m_1, \dotsc, m_n), m_i \in \{ 0, 1 \} \) which specifies the acquired spatial frequencies; note that in this case $y\in \mathbb{C}^n$.
As we need it for the inference algorithms, we denote with $A^\ast = \mathcal{F}^{-1}M$ the adjoint of \( A \).
For \gls{ct} \(A : \mathbb{R}^n \to \mathbb{R}^n : x \mapsto M\mathcal{R}x \), consists of the Radon transform \( \mathcal{R} \) and another binary mask $M$.
As before, we denote the adjoint of the operator \(A\) with \(A^\ast = \mathcal{R}^{-1}M\).
Since our simulated problems are noise-free, a classical variational reconstruction approach amounts to finding
\begin{equation}
    \argmin_x R(x) \text{ subject to } Ax = y,
    \label{eq:constrained optimization problem}
\end{equation}
where $R : \mathbb{R}^n \to \mathbb{R}$ is a regularizer. 
However, for some experiments we want to trade off the strength of the regularizer and the data fidelity.
Thus we relax the constraint and instead consider the problem of finding
\begin{equation}
    \argmin_x\, \frac{\lambda}{2}\norm{Ax - y}^2_2 + R(x),
    \label{eq:relaxed optimization problem}
\end{equation}
where $\lambda \geq 0$. 

From a Bayesian perspective,~\cref{eq:relaxed optimization problem} is the maximum a-posteriori estimate given a prior \( p_{X_0} \propto \exp(-R\bigl(\,\cdot\,)\bigr) \).
Diffusion models have recently shown remarkable results in modeling---or at least drawing samples from---\( p_{X_0} \).
In the following sections, we briefly review diffusion models and how they can be used to act as a regularizer in inverse problems of the form~\cref{eq:relaxed optimization problem}.

\subsection{Diffusion models}
Score-based diffusion models introduced by Song \etal \cite{Song2021Score} consider a family of random variables \( X_t \) indexed by a continuous variable \( t \in [0, T] \).
Here, \( X_0 \) is the random variable corresponding to the reference data whose density \( p_{X_0} \) we wish to learn and from which we have a dataset of independent and identically distributed samples.
\( X_T \) follows a distribution which we can efficiently sample from---for example a normal distribution.
The transition is modeled with the \gls{sde}
\begin{equation}
    \mathrm{d}X = f(X, t)\mathrm{d}t + g(t) \mathrm{d}w,
    \label{SDE}
\end{equation}
where $f : \mathbb{R}^n \times [0, T] \rightarrow \mathbb{R}^n$ is the \emph{drift coefficient}, $g : [0, T] \rightarrow \mathbb{R}$ is the \emph{diffusion coefficient}, and \( w \) denotes the standard Wiener process.
We choose $f$ and $g$ according to the variance exploding variant in \cite{Song2021Score}.

The reverse process of \cref{SDE} can be described by the \gls{sde}~\cite{Anderson1982}
\begin{equation}
	\mathrm{d}X = \bigl[ f(X, t) - g^2(t) \nabla \log p_{X_t}(X) \bigr] \mathrm{d}t + g(t) \mathrm{d} \tilde{w},\label{eq:rev_sde}
\end{equation}
where $\tilde{w}$ is a reverse Wiener process.
In general, the score \( \nabla \log p_{X_t} \) is not tractable, thus the main idea is to estimate it with a time-conditioned neural network \( s_{\theta} : \mathbb{R}^n \times [0, T] \to \mathbb{R}^n \) such that $s_{\theta}(\,\cdot\,, t) \approx \nabla\log p_{X_t}$, which can be trained using score matching~\cite{Hyvarinen2005,Song2020}:
The optimal parameters are found by
\begin{equation}
	\min_\theta \int_0^T \sigma^2(t) \mathbb{E}_{X_0, X_t\mid X_0} \Bigl[ \norm*{s_{\theta}(X_t, t)- \nabla_1 \log p_{X_t \mid X_0} (X_t | X_0) }_2^2 \Bigr] \mathrm{d}t.\label{eq:score matching}
\end{equation}

\subsection{Application to Medical Image Reconstruction}
There are two main approaches to generating unconditional samples from score-based diffusion models: \gls{ald} and the \gls{pc} algorithm introduced in \cite{Song2021Score}. The latter alternates between predictor steps---which numerically solve the reverse \gls{sde} \cref{eq:rev_sde}---and corrector steps---which are Langevin iterations themselves. We use both methods with extensions to allow for conditional sampling: the modification to the former was introduced by Jalal \etal \cite{Jalal2021}, the one to the latter by Chung and Ye \cite{Chung2022Score}.
They both work by imposing a data-consistency mapping of the form 
\begin{equation}
	x_i \leftarrow \mathfrak{Re}\bigl( x_i + \lambda A^\ast (y - A x_i) \bigr),\label{data_consistency}
\end{equation}
where the parameter \( \lambda\in [0,1] \) controls how heavily it is weighted; \( \lambda = 1 \) unless it is stated otherwise.
Note that \( \lambda = 0 \) corresponds to unconditional sampling.
A detailed description of the algorithms alongside the corresponding pseudo-code can be found in~\cref{sec:algo}; due to resource limitations, we use \gls{ald} sampling only on \gls{mri} data.

\subsection{Network Architecture}
The original score model \cite{Song2021Score} is a heavily modified U-Net~\cite{Ronneberger2015} with a self-attention module~\cite{Vaswani2017,Wang2018} at the lowest resolution.
Let us now introduce the notion of the depth $d$ of the network \( s_\theta \), which corresponds to the number of resolutions it operates on.
We use five models altogether:
the original model with a depth of 4 and a self-attention block; we denote this by $d=4*$; one where only the self-attention block is missing ($d=4$); and three more with $d = 3, 2, 1$.

The residual blocks $RB$ follow the design used by BigGAN~\cite{Brock2019}, the up- and down-sampling blocks ($\mathrm{Up}$ and $\mathrm{Down}$) feature an additional convolution with a binomial filter of order 3 at the higher resolution.~\cite{Zhang2019}
Time conditioning is performed with Gaussian Fourier projection~\cite{Tancik2020}, the values are added as a bias in each residual block.
The feature maps have 128 channels at the input and output resolution of $320\times 320$ and 256 channels at all other resolutions.
With this we can define three blocks, the encoder block $E^i$, the latent space block $L$ and the decoder block $D^i$ at depth \( i \):
\begin{align}
	E^i &= \mathrm{Down}^i \circ \mathit{RB}^i_4 \circ \cdots \circ \mathit{RB}^i_1,\\
	L &= \mathit{RB}_{11}^L(\cdot, a_{\mathrm{Down}_1^{L-1}}) \circ \mathit{RB}_{10}^L(\cdot, a_{\mathit{RB}_1^L}) \circ \cdots \circ \mathit{RB}_7^L(\cdot, a_{\mathit{RB}_4^L}) \circ \mathit{RB}_6^L\\
	&\quad \circ \mathrm{attention} \circ \mathit{RB}_5^L \circ \mathit{RB}_4^L \circ \cdots \circ \mathit{RB}_1^L \notag\\
	D^i &= \mathit{RB}^i_9(\cdot, a_{\mathit{RB}^i_1}) \circ \cdots \circ \mathit{RB}^i_6(\cdot, a_{\mathit{RB}^i_4}) \circ \mathit{RB}^i_5(\cdot, a_{\mathrm{Down}^{i-1}_1}) \circ \: \mathrm{Up}^i;
\end{align}
where the attention block is only present if $d=4*$.
Then we can describe the entire network depending on its depth $d$ by
\begin{equation}
	s_\theta = \mathrm{Conv1x1} \circ \: \mathrm{Conv3x3} \circ D^{d-3} \circ \cdots \circ D^{d} \circ L \circ E^{d} \circ \cdots \circ E^{d-3} \circ \: \mathrm{Conv3x3}.
\end{equation}
The additional $3\times 3$ convolutions transform the input from 1 to 128 channels and the output from 128 to 1.
For the sake of simplicity, we omit normalization blocks as well as the time conditioning.
If the index of an encoder or decoder block is less than or equal to zero, it is omitted.
Note that this is a U-Net style architecture and hence the input to the $\mathit{RB}$s of the decoders is concatenated with the activations $a_{\mathit{RB}_j^i}$ of the respective residual blocks of the encoder.

\subsubsection{Calculating the Receptive Field Sizes}
To be able to understand our results, it is important to know the size of the network's receptive field dependent on the depth \( d \).
The receptive field is defined as all the pixels in the input image of a network that contribute to a specific pixel of the output image \cite{Le2018};
we will denote it by $r_0$, where $r_0\times r_0$ is the size of the respective image patch.
Araujo \etal \cite{Araujo2019} have derived the recurrence relation
\begin{equation}
    r_{l-1} = s_l \cdot r_l + (k_l - s_l),
\end{equation}
for the receptive field size, where $s_l$ and $k_l$ denote the stride and kernel size of layer $l \in 1,\dots, L$, respectively, and $r_l$ denotes the receptive field in the final output image with respect to the feature map $f_l$. 

The receptive field sizes are illustrated as red squares in \cref{uncond_sampl}---exact numerical results can be found in \cref{tab:r0}. It is important that only once \(r_0 > 640\) is it possible that all pixels of the image can communicate with each other.

\begin{table}
	\centering
	\caption{The receptive field size \( r_0 \) dependent on the network's depth \( d \); if \( r_0 > 640 \), all pixels can communicate with each other.}\label{tab:r0}
	\begin{tabular}{c*{5}{S[table-format=>3]}}\toprule
		& {\( d=4* \)} & {\( d=4 \)} & {\( d=3 \)} & {\( d=2 \)} & {\( d=1 \)}\\\midrule
		\( r_0 \) & >640 & >640 & 331 & 143 & 49 \\\bottomrule
	\end{tabular}
\end{table}

Note, however, that Luo \etal~\cite{Luo2016} have shown that the theoretical receptive field typically overestimates the effective receptive field---which they define as the area of pixels that actually meaningfully contribute---significantly.
As an example, for $d=1$, the output is influenced by significantly less than $49\times 49$ pixels.
In addition, the influence of input pixels on output pixels decreases with their radial distance; the profile resembles a Gaussian.

\subsection{Training}
We use PyTorch~\cite{Paszke2019} to implement our experiments, the code is available on Github.\footnote{\url{https://github.com/VLOGroup/bigger-isnt-always-better}} We train all five models on \gls{mri}, \gls{ct} and natural image data. More specifically, the \gls{mri} data consists of the supplied reconstructions of the coronal proton-density weighted knee (CORPD) split of the fastMRI~\cite{Knoll2020} dataset, where we use the 15 central slices of each sample, resulting in a total of \num{7260} training slices. For CT data we use the 120 central slices of 250 samples from the LoDoPaB-CT \cite{Leuschner2021} data set which features thoracic scans, resulting in \num{30000} training samples. As natural images we use the CelebA-HQ data set~\cite{Karras2018} consisting of \num{30000} images. The latter two data sets have larger images than fastMRI and we thus have to  downsample to  $320 \times 320$ pixels; furthermore, since CelebA-HQ features color images, we transform them to grayscale.

We optimize~\cref{eq:score matching} for \num{2500000} iterations with Adam~\cite{Kingma2017}, using $\beta_1 = \num{0.9}$ and $\beta_2 = \num{0.999}$, a linear warmup schedule for the first \num{5000} iterations with a final value of \num{2e-4}, gradient clipping~\cite{Arjovsky2017} at 1 and an exponential moving average~\cite{Tarvainen2017} with a rate of \num{0.999}.

\subsection{Evaluation}
We evaluate our models both on in- and out-of-distribution data. In the context of \gls{mri}, the CORPD split serves as in-distribution data, while the fat-suppressed knee (CORPDFS) and brain data from fastMRI serve as OOD data. In-distrubtion CT data comes from the test split of the LoDoPaB-CT data set, while head scans from the RSNA Intracranial Hemorrhage Detection challenge serve as OOD data \cite{Hooper2021}. All test datasets consist of \num{100} slices, where the knee sets feature the central \num{10} slices of \num{10} randomly chosen volumes and the brain dataset of the first \num{5} slices of \num{20} randomly chosen volumes.

On the \gls{mri} data, we use five different undersampling masks:
4- and 8-fold 1D Gaussian and 4-fold 2D Gaussian undersampling, a radial mask with approximately 11-fold and a Poisson disk sampled mask with 15-fold acceleration.
The Gaussian 1D masks feature a fully-sampled center region which spans \qty{4}{\percent} of phase-encoding lines.
As undersampled \gls{ct} data we use sparse-view \gls{ct}, \ie only $N_\theta$ equidistantly spaced angles on the half circle $\theta \in [0, \pi ]$ are acquired---in this work we use $N_\theta \in \{ 60, 30 \} $.

To offer a comparison with other reconstruction techniques, we use \gls{tv} as a hand-crafted prior and U-Net as a supervised method for \gls{mri} only. In particular, we use Charbonnier smoothed \gls{tv} \cite{Charbonnier1999,Bruhn2005,Knoll2011} defined by
\begin{equation}
    \mathrm{TV}(x) = \sum_{i,j=1}^{320} \sqrt{(\nabla_h x)_{i, j}^2 + (\nabla_v x)_{i, j}^2 + \epsilon^2},
\end{equation}
with $\epsilon = 10^{-3}$.
We solve the optimization problem of the form \cref{eq:relaxed optimization problem} with \( R = \mathrm{TV} \) with accelerated proximal gradient descent, where the proximal map realizes the projection onto real-valued images.
We use the U-Net \cite{Ronneberger2015} implementation from fastMRI \cite{Knoll2020} with standard parameters and train on the CORPD data set using 4x Gaussian 1D undersampling in phase-encoding direction with \qty{8}{\percent} of the center lines densely sampled.
Note that we include these two methods for simple comparison only, Chung and Ye \cite{Chung2022Score} have already shown the $d=4*$ model to be superior to them.

\section{Results}
\subsection{Unconditional Sampling}
\begin{figure}
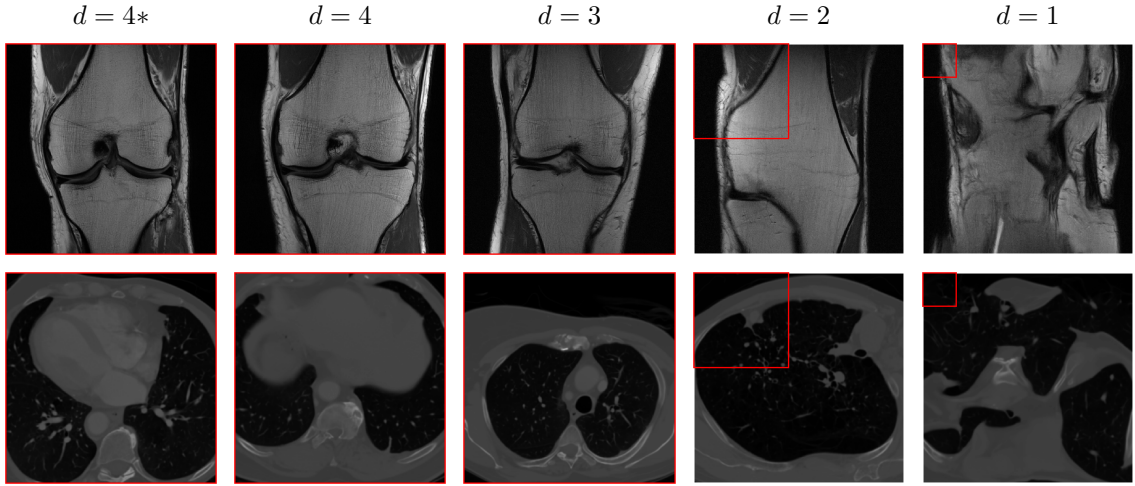

    \centering
    \def\names{{%
        "$d=4*$",%
        "$d=4$",
        "$d=3$",%
        "$d=2$",
        "$d=1$"%
    }}
    \resizebox{\linewidth}{!}{%
        \begin{tikzpicture}
            \foreach \model [count=\modeli] in {%
                fastmri_knee_4_attention_sample_0.png,
                fastmri_knee_4_sample_0.png,
                fastmri_knee_3_sample_0.png,
                fastmri_knee_2_sample_0.png,
                fastmri_knee_1_sample_0.png
            } {%
                \node (\modeli) at (3 * \modeli, 0) {\includegraphics[width=3cm]{figures/\model}};
                \node[above=0.01cm of \modeli, inner sep = 0pt] {\pgfmathparse{\names[\modeli-1]}\pgfmathresult};
            }%
            \foreach \model [count=\modeli] in {%
                ct_4_attention_sample_0.png,
                ct_4_sample_0.png,
                ct_3_sample_0.png,
                ct_2_sample_0.png,
                ct_1_sample_0.png
            } {%
                \node (\modeli) at (3 * \modeli, -3) {\includegraphics[width=3cm]{figures/\model}};
            }%
        \end{tikzpicture}
    }%
    \vspace{-5mm}\caption{Unconditional samples generated using the PC algorithm; the red squares indicate the size of the receptive field of its center pixel.}\label{uncond_sampl}
\end{figure}

Before analyzing the reconstruction performance, we consider the unconditional samples the networks generate. \Cref{uncond_sampl} shows both \gls{mri} and \gls{ct} samples for each network, drawn with the predictor--corrector algorithm.

We can see that the three largest networks can adequately represent the global structures of a knee \gls{mri} and thorax \gls{ct}, respectively.
Both networks with $d=1, 2$ still show valid local structures present in \gls{mri} images whose size approximately corresponds to the receptive field;
however, they fail to resolve global structures.
In case of \gls{mri}, the local model can locate the knee due to the imposed zero-boundary conditions of the convolutions.

\subsection{Reconstruction Performance}
\begin{figure}[ht]
	\centering
	\resizebox{\linewidth}{!}{
	\begin{tabular}{rccp{3em}cl}
		\multicolumn{3}{c}{\Huge FastMRI-trained} & & \multicolumn{2}{c}{\Huge CT-trained}\\
		\multicolumn{6}{c}{\vspace{0.02cm}} \\
		\begin{tikzpicture}[baseline,trim axis left]
			\begin{axis}[cycle list name=mycolorsmark, thick, xtick = {0, 1, 2, 3, 4}, xmin=-0.5, xmax=4.5, xticklabels={$d=4*$, $d=4$, $d=3$, $d=2$, $d=1$}, ymin=28, ymax=37, title=\LARGE CORPD (ID), ylabel=\Large PSNR in \unit{\decibel}, legend columns=-1, legend entries={G-1D x4, G-1D x8, G-2D x4, R x11, P x15}, legend to name=named, legend style={thick}, xticklabel=\empty, ticklabel style = {font=\large}]
				\foreach \i in {1, 2, ..., 5}{
					\addplot table[x index = 0, y index = \i, col sep = comma] {./data/mean_psnr_corpd.csv};
				}
				\foreach \i in {1, 2, ..., 5}{
					\addplot table[x index = 0, y index = \i, col sep = comma] {./data/jalal_mean_psnr_corpd.csv};
				}
			\end{axis}
		\end{tikzpicture}  &
		\begin{tikzpicture}[baseline]
			\begin{axis}[cycle list name=mycolorsmark, thick, xtick = {0, 1, 2, 3, 4}, xmin=-0.5, xmax=4.5, xticklabels={$d=4*$, $d=4$, $d=3$, $d=2$, $d=1$}, ymin=28, ymax=37, yticklabel=\empty, title=\LARGE CORPDFS (OOD), xticklabel=\empty]
				\foreach \i in {1, 2, ..., 5}{
					\addplot table[x index = 0, y index = \i, col sep = comma] {./data/mean_psnr_corpdfs.csv};
				}
				\foreach \i in {1, 2, ..., 5}{
					\addplot table[x index = 0, y index = \i, col sep = comma] {./data/jalal_mean_psnr_corpdfs.csv};
				}
			\end{axis}
		\end{tikzpicture} & 
		\begin{tikzpicture}[baseline]
			\begin{axis}[cycle list name=mycolorsmark, thick, xtick = {0, 1, 2, 3, 4}, xmin=-0.5, xmax=4.5, xticklabels={$d=4*$, $d=4$, $d=3$, $d=2$, $d=1$}, ymin=28, ymax=37, yticklabel=\empty, title=\LARGE Brain (OOD), xticklabel=\empty]
				\foreach \i in {1, 2, ..., 5}{
					\addplot table[x index = 0, y index = \i, col sep = comma] {./data/mean_psnr_brain.csv};
				}
				\foreach \i in {1, 2, ..., 5}{
					\addplot table[x index = 0, y index = \i, col sep = comma] {./data/jalal_mean_psnr_brain.csv};
				}
			\end{axis}
		\end{tikzpicture} & & 
		\begin{tikzpicture}[baseline]
			\begin{axis}[cycle list name=mycolorsmark, thick, xtick = {0, 1, 2, 3, 4}, xmin=-0.5, xmax=4.5, xticklabels={$d=4*$, $d=4$, $d=3$, $d=2$, $d=1$}, ymin=28, ymax=37, yticklabel=\empty, title=\LARGE Thorax CT (ID), legend columns=-1, legend entries={$N_\theta = 60$, $N_\theta=30$}, legend to name=named_ct, legend style={thick}, xticklabel=\empty]
				\foreach \i in {1, 2}{
					\addplot table[x index = 0, y index = \i, col sep = comma] {./data/mean_psnr_ct.csv};
				}
			\end{axis}
		\end{tikzpicture} & 
		\begin{tikzpicture}[baseline,trim axis right]
			\begin{axis}[cycle list name=mycolorsmark, thick, xtick = {0, 1, 2, 3, 4}, xmin=-0.5, xmax=4.5, xticklabels={$d=4*$, $d=4$, $d=3$, $d=2$, $d=1$}, ymin=28, ymax=37, yticklabel=\empty, title=\LARGE Head CT (OOD), xticklabel=\empty]
				\foreach \i in {1, 2}{
					\addplot table[x index = 0, y index = \i, col sep = comma] {./data/mean_psnr_ct_head.csv};
				}
			\end{axis}
		\end{tikzpicture} \\ 
		\multicolumn{5}{c}{\vspace{0.075cm}} \\
		\begin{tikzpicture}[baseline,trim axis left]
			\begin{axis}[cycle list name=mycolorsmark, thick, xtick = {0, 1, 2, 3, 4}, xmin=-0.5, xmax=4.5, xticklabels={$d=4*$, $d=4$, $d=3$, $d=2$, $d=1$}, ymin=0.5, ymax=0.9, ylabel=\Large SSIM in \unit{\arbitraryunit}, ticklabel style = {font=\large}]
				\foreach \i in {1, 2, ..., 5}{
					\addplot table[x index = 0, y index = \i, col sep = comma] {./data/mean_ssim_corpd.csv};
				}
				\foreach \i in {1, 2, ..., 5}{
					\addplot table[x index = 0, y index = \i, col sep = comma] {./data/jalal_mean_ssim_corpd.csv};
				}
			\end{axis}
		\end{tikzpicture} &
		\begin{tikzpicture}[baseline]
			\begin{axis}[cycle list name=mycolorsmark, thick, xtick = {0, 1, 2, 3, 4}, xmin=-0.5, xmax=4.5, xticklabels={$d=4*$, $d=4$, $d=3$, $d=2$, $d=1$}, ymin=0.5, ymax=0.9, yticklabel=\empty, xticklabel style = {font=\large}]
				\foreach \i in {1, 2, ..., 5}{
					\addplot table[x index = 0, y index = \i, col sep = comma] {./data/mean_ssim_corpdfs.csv};
				}
				\foreach \i in {1, 2, ..., 5}{
					\addplot table[x index = 0, y index = \i, col sep = comma] {./data/jalal_mean_ssim_corpdfs.csv};
				}
			\end{axis}
		\end{tikzpicture} &
		\begin{tikzpicture}[baseline]
			\begin{axis}[cycle list name=mycolorsmark, thick, xtick = {0, 1, 2, 3, 4}, xmin=-0.5, xmax=4.5, xticklabels={$d=4*$, $d=4$, $d=3$, $d=2$, $d=1$}, ymin=0.5, ymax=0.9, yticklabel=\empty, xticklabel style = {font=\large}]
				\foreach \i in {1, 2, ..., 5}{
					\addplot table[x index = 0, y index = \i, col sep = comma] {./data/mean_ssim_brain.csv};
				}
				\foreach \i in {1, 2, ..., 5}{
					\addplot table[x index = 0, y index = \i, col sep = comma] {./data/jalal_mean_ssim_brain.csv};
				}
			\end{axis}
		\end{tikzpicture} & &
		\begin{tikzpicture}[baseline]
			\begin{axis}[cycle list name=mycolorsmark, thick, xtick = {0, 1, 2, 3, 4}, xmin=-0.5, xmax=4.5, xticklabels={$d=4*$, $d=4$, $d=3$, $d=2$, $d=1$}, ymin=0.5, ymax=0.9, yticklabel=\empty, xticklabel style = {font=\large}]
				\foreach \i in {1, 2}{
					\addplot table[x index = 0, y index = \i, col sep = comma] {./data/mean_ssim_ct.csv};
				}
			\end{axis}
		\end{tikzpicture} &
		\begin{tikzpicture}[baseline,trim axis right]
			\begin{axis}[cycle list name=mycolorsmark, thick, xtick = {0, 1, 2, 3, 4}, xmin=-0.5, xmax=4.5, xticklabels={$d=4*$, $d=4$, $d=3$, $d=2$, $d=1$}, ymin=0.5, ymax=0.9, yticklabel=\empty, xticklabel style = {font=\large}]
				\foreach \i in {1, 2}{
					\addplot table[x index = 0, y index = \i, col sep = comma] {./data/mean_ssim_ct_head.csv};
				}
			\end{axis}
		\end{tikzpicture} \\
		\multicolumn{6}{c}{\vspace{0.75cm}} \\
		\multicolumn{6}{c}{\Huge CelebA-trained}\\
		\multicolumn{6}{c}{\vspace{0.02cm}} \\
	\begin{tikzpicture}[baseline,trim axis left]
		\begin{axis}[cycle list name=mycolorsmark, thick, xtick = {0, 1, 2, 3, 4}, xmin=-0.5, xmax=4.5, xticklabels={$d=4*$, $d=4$, $d=3$, $d=2$, $d=1$}, ymin=-2.5, ymax=2.5, ylabel=\Large $\Delta$ PSNR in \unit{\decibel}, title=\LARGE CORPD (ID), xticklabel=\empty, ticklabel style = {font=\large}]
			\pgfplotsset{cycle list shift=-1}
			\addplot[mark=none, black, samples=2] {0};
			\foreach \i in {1, 2, ..., 5}{
				\addplot table[x index = 0, y index = \i, col sep = comma] {./data/mean_psnr_corpd_celeba.csv};
			}
			\foreach \i in {1, 2, ..., 5}{
				\addplot table[x index = 0, y index = \i, col sep = comma] {./data/jalal_mean_psnr_corpd_celeba.csv};
			}
		\end{axis}
	\end{tikzpicture}  &
	\begin{tikzpicture}[baseline]
	\begin{axis}[cycle list name=mycolorsmark, thick, xtick = {0, 1, 2, 3, 4}, xmin=-0.5, xmax=4.5, xticklabels={$d=4*$, $d=4$, $d=3$, $d=2$, $d=1$}, ymin=-2.5, ymax=2.5, yticklabel=\empty, title=\LARGE CORPDFS (OOD), xticklabel=\empty]
		\pgfplotsset{cycle list shift=-1}
		\addplot[mark=none, black, samples=2] {0};
		\foreach \i in {1, 2, ..., 5}{
			\addplot table[x index = 0, y index = \i, col sep = comma] {./data/mean_psnr_corpdfs_celeba.csv};
		}
		\foreach \i in {1, 2, ..., 5}{
			\addplot table[x index = 0, y index = \i, col sep = comma] {./data/jalal_mean_psnr_corpdfs_celeba.csv};
		}
	\end{axis}
	\end{tikzpicture} & 
	\begin{tikzpicture}[baseline]
	\begin{axis}[cycle list name=mycolorsmark, thick, xtick = {0, 1, 2, 3, 4}, xmin=-0.5, xmax=4.5, xticklabels={$d=4*$, $d=4$, $d=3$, $d=2$, $d=1$}, ymin=-2.5, ymax=2.5, yticklabel=\empty, title=\LARGE Brain (OOD), xticklabel=\empty]
		\pgfplotsset{cycle list shift=-1}
		\addplot[mark=none, black, samples=2] {0};
		\foreach \i in {1, 2, ..., 5}{
			\addplot table[x index = 0, y index = \i, col sep = comma] {./data/mean_psnr_brain_celeba.csv};
		}
		\foreach \i in {1, 2, ..., 5}{
			\addplot table[x index = 0, y index = \i, col sep = comma] {./data/jalal_mean_psnr_brain_celeba.csv};
		}
	\end{axis}
	\end{tikzpicture} & & 
	\begin{tikzpicture}[baseline]
	\begin{axis}[cycle list name=mycolorsmark, thick, xtick = {0, 1, 2, 3, 4}, xmin=-0.5, xmax=4.5, xticklabels={$d=4*$, $d=4$, $d=3$, $d=2$, $d=1$}, ymin=-2.5, ymax=2.5, yticklabel=\empty, title=\LARGE Thorax CT (ID), xticklabel=\empty]
		\pgfplotsset{cycle list shift=-1}
		\addplot[mark=none, black, samples=2] {0};
		\foreach \i in {1, 2}{
			\addplot table[x index = 0, y index = \i, col sep = comma] {./data/mean_psnr_ct_celeba.csv};
		}
	\end{axis}
	\end{tikzpicture} & 
	\begin{tikzpicture}[baseline,trim axis right]
	\begin{axis}[cycle list name=mycolorsmark, thick, xtick = {0, 1, 2, 3, 4}, xmin=-0.5, xmax=4.5, xticklabels={$d=4*$, $d=4$, $d=3$, $d=2$, $d=1$}, ymin=-2.5, ymax=2.5, yticklabel=\empty, title=\LARGE Head CT (OOD), xticklabel=\empty]
		\pgfplotsset{cycle list shift=-1}
		\addplot[mark=none, black, samples=2] {0};
		\foreach \i in {1, 2}{
			\addplot table[x index = 0, y index = \i, col sep = comma] {./data/mean_psnr_ct_head_celeba.csv};
		}
	\end{axis}
	\end{tikzpicture} \\ 
	\begin{tikzpicture}[baseline,trim axis left]
	\begin{axis}[cycle list name=mycolorsmark, thick, xtick = {0, 1, 2, 3, 4}, xmin=-0.5, xmax=4.5, xticklabels={$d=4*$, $d=4$, $d=3$, $d=2$, $d=1$}, ymin=-0.08, ymax=0.08, ylabel=\Large$\Delta$ SSIM in \unit{\arbitraryunit}, ticklabel style = {font=\large}]
		\pgfplotsset{cycle list shift=-1}
		\addplot[mark=none, black, samples=2] {0};
		\foreach \i in {1, 2, ..., 5}{
			\addplot table[x index = 0, y index = \i, col sep = comma] {./data/mean_ssim_corpd_celeba.csv};
		}
		\foreach \i in {1, 2, ..., 5}{
			\addplot table[x index = 0, y index = \i, col sep = comma] {./data/jalal_mean_ssim_corpd_celeba.csv};
		}
	\end{axis}
	\end{tikzpicture} &
	\begin{tikzpicture}[baseline]
	\begin{axis}[cycle list name=mycolorsmark, thick, xtick = {0, 1, 2, 3, 4}, xmin=-0.5, xmax=4.5, xticklabels={$d=4*$, $d=4$, $d=3$, $d=2$, $d=1$}, ymin=-0.08, ymax=0.08, yticklabel=\empty, scaled ticks=false, xticklabel style = {font=\large}] 
		\addplot[mark=none, black, samples=2] {0};
		\pgfplotsset{cycle list shift=-1}
		\foreach \i in {1, 2, ..., 5}{
			\addplot table[x index = 0, y index = \i, col sep = comma] {./data/mean_ssim_corpdfs_celeba.csv};
		}
		\foreach \i in {1, 2, ..., 5}{
			\addplot table[x index = 0, y index = \i, col sep = comma] {./data/jalal_mean_ssim_corpdfs_celeba.csv};
		}
	\end{axis}
	\end{tikzpicture} &
	\begin{tikzpicture}[baseline]
	\begin{axis}[cycle list name=mycolorsmark, thick, xtick = {0, 1, 2, 3, 4}, xmin=-0.5, xmax=4.5, xticklabels={$d=4*$, $d=4$, $d=3$, $d=2$, $d=1$}, ymin=-0.08, ymax=0.08, yticklabel=\empty, scaled ticks=false, xticklabel style = {font=\large}]
		\addplot[mark=none, black, samples=2] {0};
		\pgfplotsset{cycle list shift=-1}
		\foreach \i in {1, 2, ..., 5}{
			\addplot table[x index = 0, y index = \i, col sep = comma] {./data/mean_ssim_brain_celeba.csv};
			\ifthenelse{\i=1}{\label{chungline}}{}
		}
		\foreach \i in {1, 2,  ..., 5}{
			\addplot table[x index = 0, y index = \i, col sep = comma] {./data/jalal_mean_ssim_brain_celeba.csv};
			\ifthenelse{\i=1}{\label{jalalline}}{}
		}
	\end{axis}
	\end{tikzpicture} & & 
	\begin{tikzpicture}[baseline]
	\begin{axis}[cycle list name=mycolorsmark, thick, xtick = {0, 1, 2, 3, 4}, xmin=-0.5, xmax=4.5, xticklabels={$d=4*$, $d=4$, $d=3$, $d=2$, $d=1$}, ymin=-0.08, ymax=0.08, yticklabel=\empty, scaled ticks=false, xticklabel style = {font=\large}]
		\addplot[mark=none, black, samples=2] {0};
		\pgfplotsset{cycle list shift=-1}
		\foreach \i in {1, 2}{
			\addplot table[x index = 0, y index = \i, col sep = comma] {./data/mean_ssim_ct_celeba.csv};
		}
	\end{axis}
	\end{tikzpicture} &
	\begin{tikzpicture}[baseline,trim axis right]
	\begin{axis}[cycle list name=mycolorsmark, thick, xtick = {0, 1, 2, 3, 4}, xmin=-0.5, xmax=4.5, xticklabels={$d=4*$, $d=4$, $d=3$, $d=2$, $d=1$}, ymin=-0.08, ymax=0.08, yticklabel=\empty, scaled ticks=false, xticklabel style = {font=\large}]
		\addplot[mark=none, black, samples=2] {0};
		\pgfplotsset{cycle list shift=-1}
		\foreach \i in {1, 2}{
			\addplot table[x index = 0, y index = \i, col sep = comma] {./data/mean_ssim_ct_head_celeba.csv};
		}
	\end{axis}
	\end{tikzpicture} \\ \multicolumn{6}{c}{\vspace{0.1cm}} \\
	\multicolumn{3}{c}{\resizebox{!}{3em}{\pgfplotslegendfromname{named}}} & & \multicolumn{2}{c}{\resizebox{!}{3em}{\pgfplotslegendfromname{named_ct}}}
\end{tabular} }
\caption{Quantitative reconstruction results in terms of mean PSNR and SSIM.
		The masks (\emph{G}aussian, \emph{R}adial, \emph{P}oisson) or number of views $N_\theta$ are color coded.
      Solid lines (\ref*{chungline}) represent \gls{pc} sampling by Chung and Ye, dashed ones (\ref*{jalalline}) \gls{ald} sampling by Jalal~\etal}\label{resfig}
\end{figure}
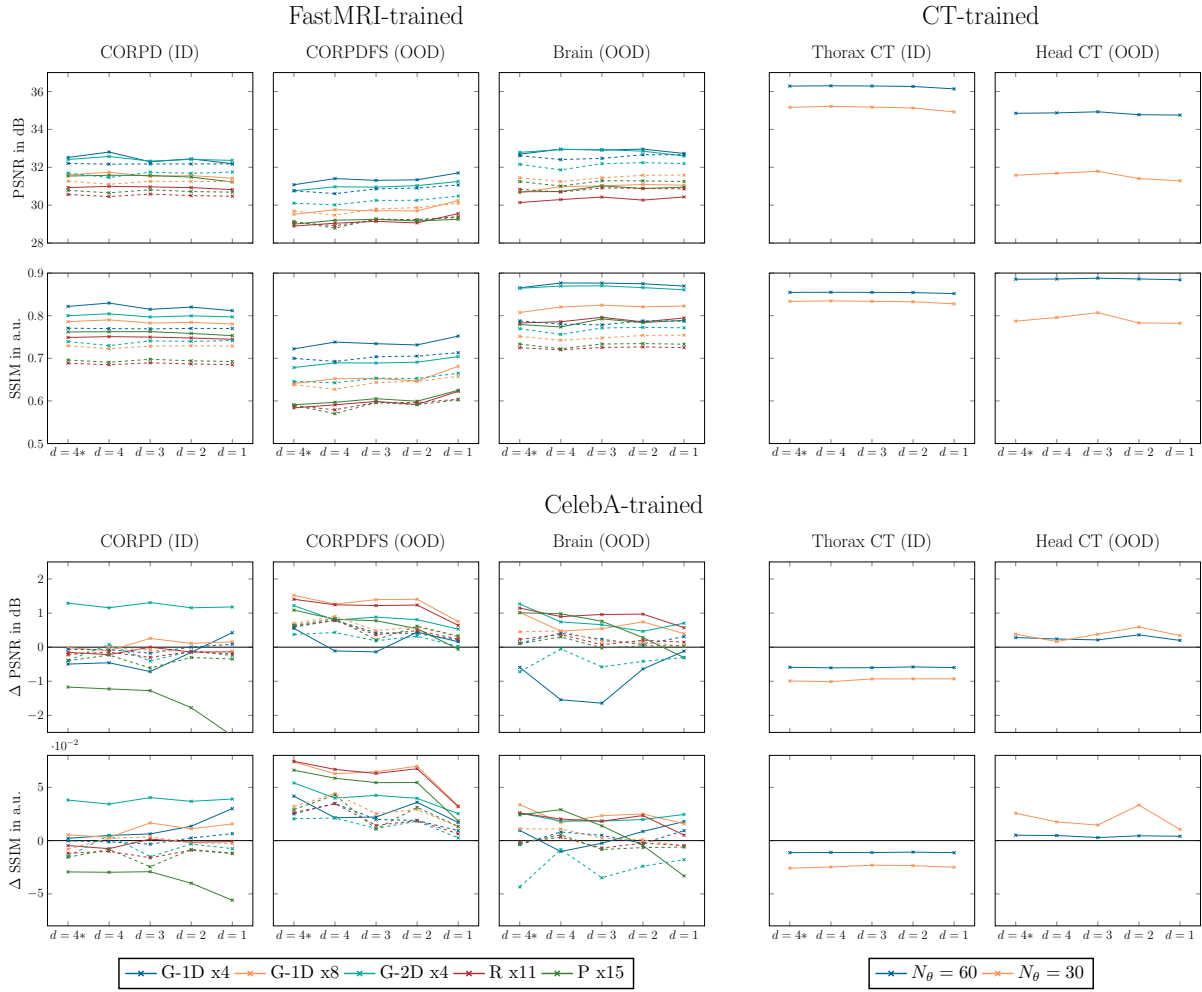

We investigate the results of the reconstructions both quantitatively and qualitatively. \Cref{resfig} shows the PSNR~\cite{Faragallah2021} and SSIM~\cite{Zhao2017} values of the conditional samples computed with the algorithms by Chung and Ye \cite{Chung2022Score} and Jalal \etal \cite{Jalal2021} for all modalities and data sets---exact values can be found in~\cref{sec:add}.
We make the following important observations regarding the models trained on the medical image datasets: 
\begin{enumerate}
    \item The standard \texttt{ncsnpp} model researchers typically use is almost never the best performing model---in particular, the attention block does not reliably increase reconstruction quality.
    \item On the CORPD and thorax CT datasets (in-distribution reconstruction), there is still a preference for the larger models (especially $d=4$) in case of the algorithm by Chung and Ye \cite{Chung2022Score}. For Jalal \etal \cite{Jalal2021}, the results are not as clear. There is only a minor difference between the performance of large and small models in both cases, particularly in the case of Jalal \etal.
    \item In out-of-distribution reconstructions, there is a preference for the smaller models.
    This is particularly pronounced in case of CORPDFS data. The advantage of the smaller models OOD is generally greater than that of the larger ones in the in-distribution case.
\end{enumerate}

The models trained on the CelebA-HQ datset consistently outperform the models trained on the medical image datasets in out-of-distribution reconstructions.
The effect is once again particularly stark in case of CORPDFS.
It is also worth noting that even on the CORPD data set, where the fastMRI-trained models work in-distribution, the CelebA-trained ones regularly perform better.

\begin{figure}
    \centering
    \def\r{_N_250_FS_no_norm/recon.png}
    \def\d{_N_250_FS_no_norm/diff.png}
    \def\lam{_lam_}
    \def\path{figures/fs_fastmri/}
    \def\pathc{figures/fs_celeba/}
    \def\maskpath{figures/fs_fastmri/mask}
    \def\maskpath{figures/fs_fastmri/mask}
    \def\underpath{figures/fs_fastmri/under}
    \def\diffpath{figures/fs_fastmri/diff_under}
    \def\tvpath{figures/fs_TV/tv}
    \def\unetpath{figures/fs_unet/unet}
    \def\png{.png}
    \def\diffpng{_diff.png}
    \def\reconpng{_recon.png}
    \def\labelfs{figures/label_fs.png}
    \def\labelbrain{figures/label_brain.png}
    \def\labelct{figures/label_ct.png}
    \def\names{{%
        "$d=4*$",%
        "$d=3$",%
        "$d=1$"%
    }}
	\def\masknames{{
		"CORPDFS Radial x11",
		"Brain Radial x11",
		"Head CT $N_\theta = 60$"%
	}}
    \resizebox{\linewidth}{!}{
    \begin{tikzpicture}
        \foreach \mask [count=\maski] in {%
            _radial,
            _poisson_AF_15,
            _gaussian2d%
        } {%
                \begin{scope}[spy using outlines={rectangle, magnification=2, size=1.48cm, connect spies}]
                \node (zfd\mask) at (0.2 - 6, -1 - 4.5*\maski) {\includegraphics[width=3cm]{\diffpath\mask\png}};
                \spy[blue] on (0.8 - 6, -0.2 - 4.5*\maski) in node[below right, semithick] at (0.2 + 0.01 - 6, -2.5 - 0.01 - 4.5*\maski);
                \end{scope}
                \begin{scope}[spy using outlines={rectangle, magnification=2, size=1.48cm, connect spies}]
                \node (zf\mask) at (0.2 - 6, -1 - 4.5*\maski) {\scalebox{1}[-1]{\includegraphics[width=3cm]{\underpath\mask\png}}};
                \ifthenelse{\maski=1}{\node[above=0.01cm of zf\mask] {\Large zero-filled};}{}
                \node [line width = 1pt, rectangle, inner sep = 0pt, color=white, draw] (mask\mask) at (0.2 - 6 - 0.9, -1 - 4.5*\maski + 0.9) {\includegraphics[width=1cm]{\maskpath\mask\png}};
                \node[left=0.01cm of zf\mask] {\rotatebox{90}{\pgfmathparse{\masknames[\maski-1]}\pgfmathresult}};
                \spy[red] on (0.8 - 6, -0.2 - 4.5*\maski) in node[below left, semithick] at (0.2 - 0.01 - 6, -2.5 - 0.01 - 4.5*\maski);
                \end{scope}
                \begin{scope}[spy using outlines={rectangle, magnification=2, size=1.48cm, connect spies}]
                \node (tvd\mask) at (0.2 - 3, -1 - 4.5*\maski) {\includegraphics[width=3cm]{\tvpath\mask\diffpng}};
                \spy[blue] on (0.8 - 3, -0.2 - 4.5*\maski) in node[below right, semithick] at (0.2 + 0.01 - 3, -2.5 - 0.01 - 4.5*\maski);
                \end{scope}
                \begin{scope}[spy using outlines={rectangle, magnification=2, size=1.48cm, connect spies}]
                \node (tv\mask) at (0.2 - 3, -1 - 4.5*\maski) {\includegraphics[width=3cm]{\tvpath\mask\reconpng}};
                \ifthenelse{\maski=1}{\node[above=0.01cm of tv\mask] {\Large TV};}{}
                \spy[red] on (0.8 - 3, -0.2 - 4.5*\maski) in node[below left, semithick] at (0.2 - 0.01 - 3, -2.5 - 0.01 - 4.5*\maski);
                \end{scope}
                
                \ifthenelse{\maski=3}{}{%
                    \begin{scope}[spy using outlines={rectangle, magnification=2, size=1.48cm, connect spies}]
                    \node (unetd\mask) at (0.2, -1 - 4.5*\maski) {\includegraphics[width=3cm]{\unetpath\mask\d}};
                    \spy[blue] on (0.8, -0.2 - 4.5*\maski) in node[below right, semithick] at (0.2 + 0.01, -2.5 - 0.01 - 4.5*\maski);
                    \end{scope}
                    \begin{scope}[spy using outlines={rectangle, magnification=2, size=1.48cm, connect spies}]
                    \node (unet\mask) at (0.2, -1 - 4.5*\maski) {\includegraphics[width=3cm]{\unetpath\mask\r}};
                    \ifthenelse{\maski=1}{\node[above=0.01cm of unet\mask] {\Large U-Net};}{}
                    \spy[red] on (0.8, -0.2 - 4.5*\maski) in node[below left, semithick] at (0.2 - 0.01, -2.5 - 0.01 - 4.5*\maski);
                    \end{scope}
                }
                
                \foreach \model [count=\modeli] in {%
                    fastmri_knee_4_attention,
                    fastmri_knee_3,
                    fastmri_knee_1%
                } { %
                        \begin{scope}[spy using outlines={rectangle, magnification=2, size=1.48cm, connect spies}]
                        \node (\model\mask) at (0.2 + 3*\modeli, -1 - 4.5*\maski) {\includegraphics[width=3cm]{\path\model\mask\d}};
                        \spy[blue] on (0.8 + 3*\modeli, -0.2 - 4.5*\maski) in node[below right, semithick] at (0.2 + 0.01 + 3*\modeli, -2.5 - 0.01 - 4.5*\maski);
                        \end{scope}
                        \begin{scope}[spy using outlines={rectangle, magnification=2, size=1.48cm, connect spies}]
                        \node (\model\mask) at (0.2 + 3*\modeli, -1 - 4.5*\maski) {\includegraphics[width=3cm]{\path\model\mask\r}};
                        \ifthenelse{\maski=1}{\node[above=0.01cm of \model\mask] {\Large\pgfmathparse{\names[\modeli-1]}\pgfmathresult};}{}
                        \spy[red] on (0.8 + 3*\modeli, -0.2 - 4.5*\maski) in node[below left, semithick] at (0.2 - 0.01 + 3*\modeli, -2.5 - 0.01 - 4.5*\maski);
                        \end{scope}
                    } %
                    \ifthenelse{\maski=2}{%
                        \begin{scope}[spy using outlines={rectangle, magnification=2, size=1.48cm, connect spies}]
                        \node (label\mask) at (0.2 + 12, -1 - 4.5*\maski) {\includegraphics[width=3cm]{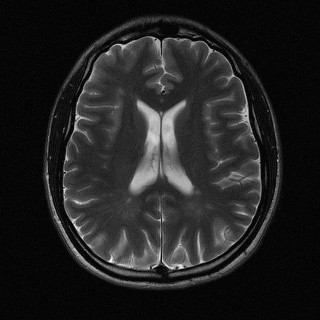}};
                        \spy[red] on (0.8 + 12, -0.2 - 4.5*\maski) in node[below left, semithick] at (0.2 - 0.01 + 12, -2.5 - 0.01 - 4.5*\maski);
                        \ifthenelse{\maski=1}{\node[above=0.01cm of label\mask] {\Large Reference};}{}
                        \end{scope}%
                    }%
                    {%
                    \ifthenelse{\maski=3}{%
                    \begin{scope}[spy using outlines={rectangle, magnification=2, size=1.48cm, connect spies}]
                    \node (label\mask) at (0.2 + 12, -1 - 4.5*\maski) {\includegraphics[width=3cm]{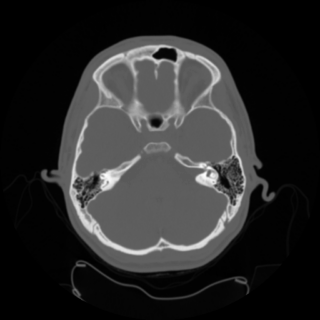}};
                    \spy[red] on (0.8 + 12, -0.2 - 4.5*\maski) in node[below left, semithick] at (0.2 - 0.01 + 12, -2.5 - 0.01 - 4.5*\maski);
                    \ifthenelse{\maski=1}{\node[above=0.01cm of label\mask] {\Large Reference};}{}
                    \end{scope}
                    }%
                    {%
                    \begin{scope}[spy using outlines={rectangle, magnification=2, size=1.48cm, connect spies}]
                    \node (label\mask) at (0.2 + 12, -1 - 4.5*\maski) {\includegraphics[width=3cm]{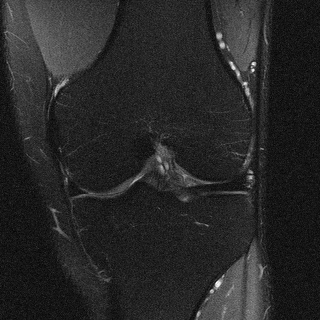}};
                    \spy[red] on (0.8 + 12, -0.2 - 4.5*\maski) in node[below left, semithick] at (0.2 - 0.01 + 12, -2.5 - 0.01 - 4.5*\maski);
                    \ifthenelse{\maski=1}{\node[above=0.01cm of label\mask] {\Large Reference};}{}
                    \end{scope}
                    }%
                    }%
                }%
                \foreach \mask [count=\maski] in {
                    _radial,
                    _poisson_AF_15,
                    _gaussian2d%
                } {%
                	
                	\begin{scope}[spy using outlines={rectangle, magnification=2, size=1.48cm, connect spies}]
                		\node (zfd\mask) at (0.2 - 6, -1 - 4.5*\maski - 14) {\includegraphics[width=3cm]{\diffpath\mask\png}};
                		\spy[blue] on (0.8 - 6, -0.2 - 4.5*\maski - 14) in node[below right, semithick] at (0.2 + 0.01 - 6, -2.5 - 0.01 - 4.5*\maski - 14);
                	\end{scope}
                	\begin{scope}[spy using outlines={rectangle, magnification=2, size=1.48cm, connect spies}]
                		\node (zf\mask) at (0.2 - 6, -1 - 4.5*\maski - 14) {\scalebox{1}[-1]{\includegraphics[width=3cm]{\underpath\mask\png}}};
                		\node [line width = 1pt, rectangle, inner sep = 0pt, color=white, draw] (mask\mask) at (0.2 - 6 - 0.9, -1 - 4.5*\maski + 0.9 - 14) {\includegraphics[width=1cm]{\maskpath\mask\png}};
                		\node[left=0.01cm of zf\mask] {\rotatebox{90}{\pgfmathparse{\masknames[\maski-1]}\pgfmathresult}};
                		\spy[red] on (0.8 - 6, -0.2 - 4.5*\maski - 14) in node[below left, semithick] at (0.2 - 0.01 - 6, -2.5 - 0.01 - 4.5*\maski - 14);
                	\end{scope}
                	
                	\begin{scope}[spy using outlines={rectangle, magnification=2, size=1.48cm, connect spies}]
                		\node (tvd\mask) at (0.2 - 3, -1 - 4.5*\maski - 14) {\includegraphics[width=3cm]{\tvpath\mask\diffpng}};
                		\spy[blue] on (0.8 - 3, -0.2 - 4.5*\maski - 14) in node[below right, semithick] at (0.2 + 0.01 - 3, -2.5 - 0.01 - 4.5*\maski - 14);
                	\end{scope}
                	\begin{scope}[spy using outlines={rectangle, magnification=2, size=1.48cm, connect spies}]
                		\node (tv\mask) at (0.2 - 3, -1 - 4.5*\maski - 14) {\includegraphics[width=3cm]{\tvpath\mask\reconpng}};
                		\spy[red] on (0.8 - 3, -0.2 - 4.5*\maski - 14) in node[below left, semithick] at (0.2 - 0.01 - 3, -2.5 - 0.01 - 4.5*\maski - 14);
                	\end{scope}
                	
                	\ifthenelse{\maski=3}{}{%
                        \begin{scope}[spy using outlines={rectangle, magnification=2, size=1.48cm, connect spies}]
                		\node (unetd\mask) at (0.2, -1 - 4.5*\maski - 14) {\includegraphics[width=3cm]{\unetpath\mask\d}};
                		\spy[blue] on (0.8, -0.2 - 4.5*\maski - 14) in node[below right, semithick] at (0.2 + 0.01, -2.5 - 0.01 - 4.5*\maski - 14);
                	\end{scope}
                	\begin{scope}[spy using outlines={rectangle, magnification=2, size=1.48cm, connect spies}]
                		\node (unet\mask) at (0.2, -1 - 4.5*\maski - 14) {\includegraphics[width=3cm]{\unetpath\mask\r}};
                		\spy[red] on (0.8, -0.2 - 4.5*\maski - 14) in node[below left, semithick] at (0.2 - 0.01, -2.5 - 0.01 - 4.5*\maski - 14);
                	\end{scope}
                 }
                	
                	\foreach \model [count=\modeli] in {%
                		celeba_4_attention, 
                		celeba_3, 
                		celeba_1%
                	} { %
                		\begin{scope}[spy using outlines={rectangle, magnification=2, size=1.48cm, connect spies}]
                			\node (\model\mask) at (0.2 + 3*\modeli, -1 - 4.5*\maski - 14) {\includegraphics[width=3cm]{\pathc\model\mask\d}};
                			\spy[blue] on (0.8 + 3*\modeli, -0.2 - 4.5*\maski - 14) in node[below right, semithick] at (0.2 + 0.01 + 3*\modeli, -2.5 - 0.01 - 4.5*\maski - 14);
                		\end{scope}
                		\begin{scope}[spy using outlines={rectangle, magnification=2, size=1.48cm, connect spies}]
                			\node (\model\mask) at (0.2 + 3*\modeli, -1 - 4.5*\maski - 14) {\includegraphics[width=3cm]{\pathc\model\mask\r}};
                			\spy[red] on (0.8 + 3*\modeli, -0.2 - 4.5*\maski - 14) in node[below left, semithick] at (0.2 - 0.01 + 3*\modeli, -2.5 - 0.01 - 4.5*\maski - 14);
                		\end{scope}
                	} %
                    \ifthenelse{\maski=2}{%
                	\begin{scope}[spy using outlines={rectangle, magnification=2, size=1.48cm, connect spies}]
                		\node (label\mask) at (0.2 + 12, -1 - 4.5*\maski - 14) {\includegraphics[width=3cm]{\labelbrain}};
                		\spy[red] on (0.8 + 12, -0.2 - 4.5*\maski - 14) in node[below left, semithick] at (0.2 - 0.01 + 12, -2.5 - 0.01 - 4.5*\maski - 14);
                	\end{scope}
                 }%
                 {%
                    \ifthenelse{\maski=3}{%
                        \begin{scope}[spy using outlines={rectangle, magnification=2, size=1.48cm, connect spies}]
                		\node (label\mask) at (0.2 + 12, -1 - 4.5*\maski - 14) {\includegraphics[width=3cm]{\labelct}};
                		\spy[red] on (0.8 + 12, -0.2 - 4.5*\maski - 14) in node[below left, semithick] at (0.2 - 0.01 + 12, -2.5 - 0.01 - 4.5*\maski - 14);
                	\end{scope}
                    }%
                    {%
                	\begin{scope}[spy using outlines={rectangle, magnification=2, size=1.48cm, connect spies}]
                		\node (label\mask) at (0.2 + 12, -1 - 4.5*\maski - 14) {\includegraphics[width=3cm]{\labelfs}};
                		\spy[red] on (0.8 + 12, -0.2 - 4.5*\maski - 14) in node[below left, semithick] at (0.2 - 0.01 + 12, -2.5 - 0.01 - 4.5*\maski - 14);
                	\end{scope}
                 }%
                 }%
                }%
    \end{tikzpicture}
    }
    \vspace{-5mm}\caption{%
        Representative reconstructions of OOD data using models trained on medical images (top) and natural ones (bottom)
        The insets show a zoom on artifacts (red border) and the corresponding magnitude of the difference between the reconstruction and the reference (blue border; 0~\protect\drawcolorbar~0.15).%
    }%
    \label{fig:fastmri reconstructions}
\end{figure}

\Cref{fig:fastmri reconstructions} offers a direct comparison of the reconstructions computed with the \gls{pc} algorithm by Chung and Ye for one representative slice from the OOD data sets. The example from the CORPDFS data demonstrates why the smallest model performs best: in all other models there are artifacts in the form of fine lines present, not so for $d=1$. These artifacts are also missing in the reconstructions from the CelebA-HQ--trained models, explaining their better performance.

\section{Discussion}
The fact that the standard \texttt{ncsnpp} model is outperformed by smaller models in almost all cases raises an important issue that has so far been overlooked in the adoption of generative approaches for MRI reconstruction. Simply downloading a model originally developed for unconditional sampling and using it as part of a posterior sampling algorithm leads to sub-optimal reconstruction performance even in the in-distribution case; it follows that it is important to adapt the model complexity to the task at hand.

The initial motivation of this work has been to increase the generalization capabilities of diffusion models, in order to make them more robust towards distribution shifts. Our results on out-of-distribution reconstruction show that we can indeed achieve this goal by decreasing the receptive field size of the models.

When developing a general prior for medical image reconstruction, there really is no in-distribution case, since one cannot reasonably make assumptions on what anatomy or contrast the model will have to deal with. The fact that the networks trained on CelebA significantly outperform those trained on medical images in the cases where they both perform out-of-distribution, shows that in order to develop such a general prior it is clearly beneficial to train on a set of natural images. We would like to stress that the same exact models are used on both \gls{mri} and \gls{ct} data and that they consistently outperform models trained on the respective medical images.

An additional advantage of this finding is that data sets containing natural images are much more readily available and significantly larger than those containing medical images. This is particularly relevant when thinking about clinical adoption, where a reconstruction algorithm has to deal with any anatomical structure, any orientation of the image plane as well as any contrast. It is unrealistic that data sets will be created for all possible modalities, given the fact that the privacy of medical data is of great concern.

\subsection{On the Effect of the Data-Consistency Mapping}
\begin{figure}
    \centering
    \resizebox{\linewidth}{!}{%
        \def\r{_CF_004_N_250_no_norm_flipped/recon_0.png}
        \def\lam{_lam_}
        \def\path{figures/lambdas/}
        \def\names{{"$d=4*$","$d=4$","$d=3$","$d=2$","$d=1$"}}
        \begin{tikzpicture}
            \foreach \m [count=\mi] in {celeba_4_attention, celeba_4, celeba_3, celeba_2, celeba_1} {%
                \node (\m) at (0, -1 - 2.8*\mi) {\includegraphics[width=2.7cm]{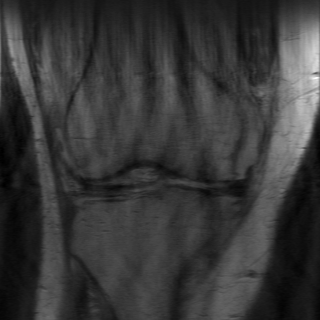}};
                \node[left=0.01cm of \m] {\rotatebox{90}{\Large\pgfmathparse{\names[\mi-1]}\pgfmathresult}};
                \ifthenelse{\mi=1}{\node[above=0.01cm of \m] {\Large zero-filled};}{}
                \foreach \l [count=\li] in {0.0, 0.001, 0.003, 0.005, 0.007, 0.009, 0.01, 0.03, 0.05, 1.0}
                    { %
                        \node (\m\li) at (0.2 + 2.8*\li, -1 - 2.8*\mi) {\includegraphics[width=2.7cm]{\path\m\lam\l\r}};
                        \ifthenelse{\mi=1}{\node[above=0.01cm of \m\li] {\Large $\lambda=\l$};}{}
                    } %
                }%
            \draw[thick] (1.5, -3) -- (1.5, -15.8);
        \end{tikzpicture}
    }%
    \vspace{-4mm}\caption{All reconstructions were computed with CelebA-trained networks using \gls{pc} sampling; $\lambda=0$ corresponds to unconditional sampling, $\lambda=1$ to the sampling algorithm used throughout the rest of this work.}\label{fig:lam}
\end{figure}

One might find it surprising that models trained on natural images are able to reconstruct undersampled medical scans with such high accuracy, better even in OOD cases, hence we decided to further investigate the parameter $\lambda$, which determines the trade-off between data-consistency and prior knowledge.
\Cref{fig:lam} shows reconstructions of a single slice from the CORPD set for different values of $\lambda$ for all CelebA-HQ--trained networks.

We can see that only for very small values of $\lambda$ (recall that by default $\lambda = 1$), the networks impart faces onto the reconstruction, while for slightly higher values ($\lambda \in [0.01, 0.05]$), we can see that there are still remnants of aliasing artifacts present. 
This shows that the data-consistency mapping is weighed very heavily in the standard implementation with $\lambda=1$, which also explains why it is not important for the score model to learn the prior distribution with a high accuracy.
Indeed, it is the reason why the CelebA-HQ-trained models can outperform the fastMRI-trained ones:
they can generalize even better, because they can learn a more diverse solution space while the strong data consistency mapping makes sure that the result is still consistent with the measurement model $A$.

\subsection{Related Work}
In~\cite{Zach2023Explicit}, the authors design networks that follow the diffusion equation by design.
In order to achieve this, they resort to one-layer networks with trainable Gaussian mixture activation function---this can be seen as an extreme case of our idea.
In the light of our findings it would be interesting to adopt their paradigm for medical image reconstruction and evaluate the generalization capabilities of the resulting model.

While preparing this paper we became aware of the recent work by Gao \etal~\cite{Gao2024}, who have also used a diffusion model trained on natural images for \gls{mri} reconstruction. However, there are several noteworthy differences, the most important one is that they use stable diffusion~\cite{Rombach2022}, which has been trained on the non-curated LAION-5B data set~\cite{Schuhmann2022}. While it contains mostly natural images, it will inadvertently also feature some \gls{mri} images; by using the curated CelebA-HQ data set, we can guarantee that the network has only ever been trained on natural images. Furthermore, we provide broader results  by  using two different algorithms and \gls{ct} as well as \gls{mri} data.

\subsection{Implications on Patched Diffusion Models}
Recently, patched diffusion models have seen a lot of research interest in a variety of inverse problems~\cite{Ozdenizci2023,Oguz2024,Ding2024}.
The fact that we have shown that the attention block does not have a positive impact in the context of medical image reconstruction has an important implication on this research area.
The idea behind these models is to make networks featuring attention agnostic to the input size.
However, if attention does not add any benefits, one could use purely convolutional networks which are size-agnostic and translation equivariant out-of-the-box.
Note that we have shown this effect only in the context of medical images; to make a more general statement more research is needed in different kinds of inverse problems.

\section{Conclusion}
We have started out with the hypothesis that diffusion-based medical image reconstruction is mainly driven by local image features. Based on this theory we have shown that a viable method to improve the generalization capability is to reduce the receptive field of the score model and to train on natural images to learn a more diverse set of local features. With these adaptions, our models can reconstruct undersampled \gls{mri} significantly better in out-of-distribution settings. Moreover, we have demonstrated that we can use the same exact models for \gls{mri} and \gls{ct} reconstruction and that they achieve better results in out-of-distribution reconstruction than models trained in the respective image domain. This marks an important step towards a general prior for medical image reconstructions.

Since our findings hold in two different settings---\gls{mri} and \gls{ct}---we believe that they will extend to other inverse problems, such as image in-painting or super-resolution, as well. However, the fact that we only consider medical image reconstruction is a limitation and we strongly encourage future research into the application in other domains.

\bibliographystyle{splncs04}
\bibliography{bibliography}

\begin{thebibliography}{10}
\providecommand{\url}[1]{\texttt{#1}}
\providecommand{\urlprefix}{URL }
\providecommand{\doi}[1]{https://doi.org/#1}

\bibitem{Aggarwal2018}
Aggarwal, H.K., Mani, M.P., Jacob, M.: Modl: Model-based deep learning architecture for inverse problems. IEEE Transactions on Medical Imaging  \textbf{38}(2),  394--405 (2019). \doi{10.1109/TMI.2018.2865356}

\bibitem{Anderson1982}
Anderson, B.D.: Reverse-time diffusion equation models. Stochastic Processes and their Applications  \textbf{12}(3),  313--326 (1982). \doi{10.1016/0304-4149(82)90051-5}

\bibitem{Araujo2019}
Araujo, A., Norris, W., Sim, J.: Computing receptive fields of convolutional neural networks. Distill  (2019). \doi{10.23915/distill.00021}, https://distill.pub/2019/computing-receptive-fields

\bibitem{Arjovsky2017}
Arjovsky, M., Chintala, S., Bottou, L.: {W}asserstein generative adversarial networks. In: Proceedings of the 34th International Conference on Machine Learning. Proceedings of Machine Learning Research, vol.~70, pp. 214--223. PMLR (06--11 Aug 2017)

\bibitem{Oguz2024}
Behrendt, F., Bhattacharya, D., Kr\"uger, J., Opfer, R., Schlaefer, A.: Patched diffusion models for unsupervised anomaly detection in brain mri. In: Oguz, I., Noble, J., Li, X., Styner, M., Baumgartner, C., Rusu, M., Heinmann, T., Kontos, D., Landman, B., Dawant, B. (eds.) Medical Imaging with Deep Learning. Proceedings of Machine Learning Research, vol.~227, pp. 1019--1032. PMLR (10--12 Jul 2024)

\bibitem{Brock2019}
Brock, A., Donahue, J., Simonyan, K.: Large scale {GAN} training for high fidelity natural image synthesis. In: International Conference on Learning Representations (2019)

\bibitem{Bruhn2005}
Bruhn, A., Weickert, J., Schn{\"o}rr, C.: Lucas/kanade meets horn/schunck: Combining local and global optic flow methods. International Journal of Computer Vision  \textbf{61}(3),  211--231 (2005). \doi{10.1023/B:VISI.0000045324.43199.43}

\bibitem{Chambolle2004}
Chambolle, A.: An algorithm for total variation minimization and applications. Journal of Mathematical Imaging and Vision  \textbf{20}(1),  89--97 (Jan 2004). \doi{10.1023/B:JMIV.0000011325.36760.1e}

\bibitem{Chang2023}
Chang, Z., Koulieris, G.A., Shum, H.P.H.: On the design fundamentals of diffusion models: A survey (2023). \doi{10.48550/arXiv.2306.04542}

\bibitem{Charbonnier1999}
Charbonnier, P., Blanc-Feraud, L., Aubert, G., Barlaud, M.: Deterministic edge-preserving regularization in computed imaging. IEEE Transactions on Image Processing  \textbf{6}(2),  298--311 (1997). \doi{10.1109/83.551699}

\bibitem{Chung2023}
Chung, H., Kim, J., Mccann, M.T., Klasky, M.L., Ye, J.C.: Diffusion posterior sampling for general noisy inverse problems. In: The Eleventh International Conference on Learning Representations (2023)

\bibitem{Chung2022Come}
Chung, H., Sim, B., Ye, J.C.: Come-closer-diffuse-faster: Accelerating conditional diffusion models for inverse problems through stochastic contraction. In: 2022 IEEE/CVF Conference on Computer Vision and Pattern Recognition (CVPR). pp. 12403--12412 (2022). \doi{10.1109/CVPR52688.2022.01209}

\bibitem{Chung2022Score}
Chung, H., Ye, J.C.: Score-based diffusion models for accelerated mri. Medical Image Analysis  \textbf{80},  102479 (2022). \doi{10.1016/j.media.2022.102479}

\bibitem{Ding2024}
Ding, Z., Zhang, M., Wu, J., Tu, Z.: Patched denoising diffusion models for high-resolution image synthesis. In: The Twelfth International Conference on Learning Representations (2024)

\bibitem{Faragallah2021}
Faragallah, O.S., El-Hoseny, H., El-Shafai, W., El-Rahman, W.A., El-Sayed, H.S., El-Rabaie, E.S.M., El-Samie, F.E.A., Geweid, G.G.N.: A comprehensive survey analysis for present solutions of medical image fusion and future directions. IEEE Access  \textbf{9},  11358--11371 (2021). \doi{10.1109/ACCESS.2020.3048315}

\bibitem{Gao2023}
Gao, S., Liu, X., Zeng, B., Xu, S., Li, Y., Luo, X., Liu, J., Zhen, X., Zhang, B.: Implicit diffusion models for continuous super-resolution. In: Proceedings of the IEEE/CVF Conference on Computer Vision and Pattern Recognition (CVPR). pp. 10021--10030 (June 2023)

\bibitem{Gao2024}
Gao, Z., Zhou, S.K.: U$^2$mrpd: Unsupervised undersampled mri reconstruction by prompting a large latent diffusion model (2024). \doi{10.48550/arXiv.2402.10609}

\bibitem{Gupta2024}
Gupta, S., Jalal, A., Parulekar, A., Price, E., Xun, Z.: Diffusion posterior sampling is computationally intractable (2024). \doi{10.48550/arXiv.2402.12727}

\bibitem{Hammernik2018}
Hammernik, K., Klatzer, T., Kobler, E., Recht, M.P., Sodickson, D.K., Pock, T., Knoll, F.: Learning a variational network for reconstruction of accelerated mri data. Magnetic Resonance in Medicine  \textbf{79}(6),  3055--3071 (2018). \doi{10.1002/mrm.26977}

\bibitem{Ho2020}
Ho, J., Jain, A., Abbeel, P.: Denoising diffusion probabilistic models. In: Larochelle, H., Ranzato, M., Hadsell, R., Balcan, M., Lin, H. (eds.) Advances in Neural Information Processing Systems. vol.~33, pp. 6840--6851. Curran Associates, Inc. (2020)

\bibitem{Hooper2021}
Hooper, S.M., Dunnmon, J.A., Lungren, M.P., Mastrodicasa, D., Rubin, D.L., R\'{e}, C., Wang, A., Patel, B.N.: Impact of upstream medical image processing on downstream performance of a head ct triage neural network. Radiology: Artificial Intelligence  \textbf{3}(4),  e200229 (2021). \doi{10.1148/ryai.2021200229}

\bibitem{Huang1999}
Huang, J., Mumford, D.: Statistics of natural images and models. In: Proceedings. 1999 IEEE Computer Society Conference on Computer Vision and Pattern Recognition (Cat. No PR00149). vol.~1, pp. 541--547 Vol. 1 (1999). \doi{10.1109/CVPR.1999.786990}

\bibitem{Hyvarinen2005}
Hyv{{\"a}}rinen, A.: Estimation of non-normalized statistical models by score matching. Journal of Machine Learning Research  \textbf{6}(24),  695--709 (2005)

\bibitem{Jalal2021}
Jalal, A., Arvinte, M., Daras, G., Price, E., Dimakis, A.G., Tamir, J.: Robust compressed sensing mri with deep generative priors. In: Advances in Neural Information Processing Systems. vol.~34, pp. 14938--14954. Curran Associates, Inc. (2021)

\bibitem{Karras2018}
Karras, T., Aila, T., Laine, S., Lehtinen, J.: Progressive growing of {GAN}s for improved quality, stability, and variation. In: International Conference on Learning Representations (2018)

\bibitem{Kawar2022}
Kawar, B., Elad, M., Ermon, S., Song, J.: Denoising diffusion restoration models. In: Advances in Neural Information Processing Systems. vol.~35, pp. 23593--23606. Curran Associates, Inc. (2022)

\bibitem{Kawar2021}
Kawar, B., Vaksman, G., Elad, M.: Snips: Solving noisy inverse problems stochastically. In: Advances in Neural Information Processing Systems. vol.~34, pp. 21757--21769. Curran Associates, Inc. (2021)

\bibitem{Kingma2017}
Kingma, D.P., Ba, J.: Adam: A method for stochastic optimization (2017). \doi{10.48550/arXiv.1412.6980}

\bibitem{Knoll2011}
Knoll, F., Bredies, K., Pock, T., Stollberger, R.: Second order total generalized variation (tgv) for mri. Magnetic Resonance in Medicine  \textbf{65}(2),  480--491 (2011). \doi{10.1002/mrm.22595}

\bibitem{Knoll2020}
Knoll, F., Zbontar, J., Sriram, A., Muckley, M.J., Bruno, M., Defazio, A., Parente, M., Geras, K.J., Katsnelson, J., Chandarana, H., et~al.: fastmri: A publicly available raw k-space and dicom dataset of knee images for accelerated mr image reconstruction using machine learning. Radiology: Artificial Intelligence  \textbf{2}(1),  e190007 (2020)

\bibitem{Le2018}
Le, H., Borji, A.: What are the receptive, effective receptive, and projective fields of neurons in convolutional neural networks? (2018). \doi{10.48550/arXiv.1705.07049}

\bibitem{Leuschner2021}
Leuschner, J., Schmidt, M., Baguer, D.O., Maass, P.: Lodopab-ct, a benchmark dataset for low-dose computed tomography reconstruction. Scientific Data  \textbf{8}(1), ~109 (2021). \doi{10.1038/s41597-021-00893-z}

\bibitem{Lugmayr2022}
Lugmayr, A., Danelljan, M., Romero, A., Yu, F., Timofte, R., Van~Gool, L.: Repaint: Inpainting using denoising diffusion probabilistic models. In: Proceedings of the IEEE/CVF Conference on Computer Vision and Pattern Recognition (CVPR). pp. 11461--11471 (June 2022)

\bibitem{Luo2023}
Luo, G., Wang, X., Blumenthal, M., Schilling, M., Rauf, E.H.U., Kotikalapudi, R., Focke, N., Uecker, M.: Generative image priors for mri reconstruction trained from magnitude-only images (2023). \doi{10.48550/arXiv.2308.02340}

\bibitem{Luo2016}
Luo, W., Li, Y., Urtasun, R., Zemel, R.: Understanding the effective receptive field in deep convolutional neural networks. In: Advances in Neural Information Processing Systems. vol.~29. Curran Associates, Inc. (2016)

\bibitem{Lustig2007}
Lustig, M., Donoho, D., Pauly, J.M.: Sparse mri: The application of compressed sensing for rapid mr imaging. Magnetic Resonance in Medicine  \textbf{58}(6),  1182--1195 (2007). \doi{10.1002/mrm.21391}

\bibitem{Ozdenizci2023}
\"Ozdenizci, O., Legenstein, R.: Restoring vision in adverse weather conditions with patch-based denoising diffusion models. IEEE Transactions on Pattern Analysis and Machine Intelligence  \textbf{45}(8),  10346--10357 (2023). \doi{10.1109/TPAMI.2023.3238179}

\bibitem{Paszke2019}
Paszke, A., Gross, S., Massa, F., Lerer, A., Bradbury, J., Chanan, G., Killeen, T., Lin, Z., Gimelshein, N., Antiga, L., Desmaison, A., Kopf, A., Yang, E., DeVito, Z., Raison, M., Tejani, A., Chilamkurthy, S., Steiner, B., Fang, L., Bai, J., Chintala, S.: Pytorch: An imperative style, high-performance deep learning library. In: Advances in Neural Information Processing Systems. vol.~32. Curran Associates, Inc. (2019)

\bibitem{Rombach2022}
Rombach, R., Blattmann, A., Lorenz, D., Esser, P., Ommer, B.: High-resolution image synthesis with latent diffusion models. In: Proceedings of the IEEE/CVF Conference on Computer Vision and Pattern Recognition (CVPR). pp. 10684--10695 (June 2022)

\bibitem{Ronneberger2015}
Ronneberger, O., Fischer, P., Brox, T.: U-net: Convolutional networks for biomedical image segmentation. In: Medical Image Computing and Computer-Assisted Intervention -- MICCAI 2015. pp. 234--241. Springer International Publishing, Cham (2015)

\bibitem{Schlemper2017}
Schlemper, J., Caballero, J., Hajnal, J.V., Price, A., Rueckert, D.: A deep cascade of convolutional neural networks for mr image reconstruction. In: Information Processing in Medical Imaging. pp. 647--658. Springer International Publishing, Cham (2017). \doi{10.1007/978-3-319-59050-9\_51}

\bibitem{Schuhmann2022}
Schuhmann, C., Beaumont, R., Vencu, R., Gordon, C., Wightman, R., Cherti, M., Coombes, T., Katta, A., Mullis, C., Wortsman, M., Schramowski, P., Kundurthy, S., Crowson, K., Schmidt, L., Kaczmarczyk, R., Jitsev, J.: Laion-5b: An open large-scale dataset for training next generation image-text models. In: Advances in Neural Information Processing Systems. vol.~35, pp. 25278--25294. Curran Associates, Inc. (2022)

\bibitem{Song2021DDIM}
Song, J., Meng, C., Ermon, S.: Denoising diffusion implicit models. In: International Conference on Learning Representations (2021)

\bibitem{Song2020}
Song, Y., Garg, S., Shi, J., Ermon, S.: Sliced score matching: A scalable approach to density and score estimation. In: Proceedings of The 35th Uncertainty in Artificial Intelligence Conference. Proceedings of Machine Learning Research, vol.~115, pp. 574--584. PMLR (22--25 Jul 2020)

\bibitem{Song2022}
Song, Y., Shen, L., Xing, L., Ermon, S.: Solving inverse problems in medical imaging with score-based generative models. In: International Conference on Learning Representations (2022)

\bibitem{Song2021Score}
Song, Y., Sohl-Dickstein, J., Kingma, D.P., Kumar, A., Ermon, S., Poole, B.: Score-based generative modeling through stochastic differential equations. In: International Conference on Learning Representations (2021)

\bibitem{Tancik2020}
Tancik, M., Srinivasan, P., Mildenhall, B., Fridovich-Keil, S., Raghavan, N., Singhal, U., Ramamoorthi, R., Barron, J., Ng, R.: Fourier features let networks learn high frequency functions in low dimensional domains. In: Advances in Neural Information Processing Systems. vol.~33, pp. 7537--7547. Curran Associates, Inc. (2020)

\bibitem{Tarvainen2017}
Tarvainen, A., Valpola, H.: Mean teachers are better role models: Weight-averaged consistency targets improve semi-supervised deep learning results. In: Advances in Neural Information Processing Systems. vol.~30. Curran Associates, Inc. (2017)

\bibitem{Vaswani2017}
Vaswani, A., Shazeer, N., Parmar, N., Uszkoreit, J., Jones, L., Gomez, A.N., Kaiser, L.u., Polosukhin, I.: Attention is all you need. In: Advances in Neural Information Processing Systems. vol.~30. Curran Associates, Inc. (2017)

\bibitem{Wang2018}
Wang, X., Girshick, R., Gupta, A., He, K.: Non-local neural networks. In: Proceedings of the IEEE Conference on Computer Vision and Pattern Recognition (CVPR) (June 2018)

\bibitem{Zach2023}
Zach, M., Knoll, F., Pock, T.: Stable deep mri reconstruction using generative priors. IEEE Transactions on Medical Imaging  \textbf{42}(12),  3817--3832 (2023). \doi{10.1109/TMI.2023.3311345}

\bibitem{Zach2023Explicit}
Zach, M., Kobler, E., Chambolle, A., Pock, T.: Product of gaussian mixture diffusion models (2024). \doi{10.1007/s10851-024-01180-3}

\bibitem{Zach2021}
Zach, M., Kobler, E., Pock, T.: Computed tomography reconstruction using generative energy-based priors. In: Proc. of the OAGM Workshop. pp. 52--58 (Dec 2021)

\bibitem{Zhang2019}
Zhang, R.: Making convolutional networks shift-invariant again. In: Proceedings of the 36th International Conference on Machine Learning. Proceedings of Machine Learning Research, vol.~97, pp. 7324--7334. PMLR (09--15 Jun 2019)

\bibitem{Zhao2017}
Zhao, H., Gallo, O., Frosio, I., Kautz, J.: Loss functions for image restoration with neural networks. IEEE Transactions on Computational Imaging  \textbf{3}(1),  47--57 (2017). \doi{10.1109/TCI.2016.2644865}

\end{thebibliography}
\appendix
\section{Sampling Algorithms}
\label{sec:algo}
We use two different posterior sampling algorithms: \gls{ald} and \gls{pc} sampling. The former has been introduced by Jalal \etal \cite{Jalal2021}, it anneals both the Langevin as well as the data consistency step and applies them at the same time, \cref{alg:Langevin} shows its exact definition.

\begin{algorithm}
	\DontPrintSemicolon
	\KwIn{\( s_{\theta}, N, N_\mathrm{start}, M, \lambda, \{ \sigma_i \}_{i=0}^N, \{ \epsilon_i \}_{i=0}^{N-1}, \{ \gamma_i \}_{i=0}^{N-1} \)}
	\KwOut{\( x_0 \sim p_{X_0 \mid Y} \)}
	\( x_{N_\mathrm{start}} \sim \mathcal{N}(0, {\mathrm{Id}}) \)\;
	\For{\( i = N_\mathrm{start}-1,\dotsc,0 \)}{
		\For{\( j = M-1, \dotsc, 0 \)}{
			\( z\sim\mathcal{N}(0, {\mathrm{Id}}) \)\;
			\( x_i\leftarrow x_{i+1} + \epsilon_i \biggl( s_{\theta} (x_{i+1}, \sigma_{i+1}) - \lambda \mathfrak{Re}\Bigl(\frac{A^\ast(A x_{i+1} - y)}{\gamma_i^2}\Bigr)\biggr) + \sqrt{2\epsilon_i} z \)
		}
	}
	\caption{%
        Annealed Langevin dynamics to sample from \( p_{X_0 \mid Y} \).
    }%
    \label{alg:Langevin}
\end{algorithm}

The parameters are taken from Jalal \etal, with the exception of the noise scales $\{ \sigma_i \}_{i=0}^N$, which are populated from a geometric series between $\sigma_N=378$ and $\sigma_1 = 0.01$---where $N=500$. As in the original implementation, we do not actually start to sample at $N$, but rather at $N_\mathrm{start}$. This value is chosen in such a way that $\sigma_{N_\mathrm{start}} \approx 1$, because the reconstruction in initialized from $\mathcal{N}(0,I)$. We set $N_\mathrm{start} = 230$; note that this is a similar, albeit smaller ($\approx 2\times$), acceleration technique as introduced in \cite{Chung2022Come}.

\begin{algorithm}
	\DontPrintSemicolon
	\KwIn{\( s_{\theta}, N, \lambda, \{ \sigma_i \}_{i=0}^N, \{ \epsilon_i \}_{i=0}^{N-1} \)}
	\KwOut{\( x_0 \sim p_{X_0 \mid Y} \)}
	\( x_N \sim \mathcal{N}(0, \sigma_N^2{\mathrm{Id}}) \)\;
	\For{\( i = N-1,\dotsc,0 \)}{
		\( z\sim\mathcal{N}(0, {\mathrm{Id}}) \)\;
		\( x_i\leftarrow x_{i+1} + (\sigma_{i+1}^2 - \sigma_i^2) s_{\theta} (x_{i+1}, \sigma_{i+1}) + \sqrt{\sigma_{i+1}^2 - \sigma_i^2} z \) \tcp*{predictor}
		\( x_i \leftarrow \mathfrak{Re}\bigl( x_i - \lambda A^\ast(Ax_i - y) \bigr) \)\tcp*{data consistency}
		\( z\sim\mathcal{N}(0, {\mathrm{Id}}) \)\;
		\( x_i\leftarrow x_{i} + \epsilon_i s_{\theta} (x_i, \sigma_i) + \sqrt{2\epsilon_i} z \) \tcp*{corrector}
		\( x_i \leftarrow \mathfrak{Re}\bigl( x_i - \lambda A^\ast(Ax_i - y) \bigr) \)\tcp*{data consistency}
}
	\caption{%
        \gls{pc} algorithm to sample from \( p_{X_0 \mid Y} \).
    }%
    \label{alg:cond_PC}
\end{algorithm}

In the conditional \gls{pc} algorithm introduced by \cite{Chung2022Score} and shown in \cref{alg:cond_PC}, we have introduced an additional parameter $\lambda$, which controls the trade-off between data consistency and prior knowledge; unless noted otherwise, we set $\lambda = 1$. While Chung and Ye~\cite{Chung2022Score} use $N=\num{2000}$, we have to choose a lower value due to computational constraints and settle on $N=250$. From their study of the influence of the number of noise scales $N$, we can reasonably believe that it is large enough. The values of $\{ \epsilon_i \}_{i=0}^{N-1}$ are populated with the same values as in the paper by  Chung and Ye, the noise scales are chosen analogously to the \gls{ald} algorithm.

\section{Additional Reconstructions}
\label{sec:add}
\Cref{tab:mri,tab:ct} show the numerical values of PSNR and SSIM used to construct Figure 3. \Cref{tab:mri} concerns MRI and \cref{tab:ct} CT reconstructions. Additionally, they feature the number of trainable parameters, size of the receptive field $r_0$ and the reconstruction times for the respective models.

\robustify{\bfseries}
\robustify{\itshape}

\begin{table}
    \def\mymidrule{\cmidrule(r{.5em}){1-5}\cmidrule(lr{.5em}){6-10}\cmidrule(l{.5em}){11-15}}
    \centering
    \caption{%
        Quantitative reconstruction results in terms of PSNR and SSIM (mean $\pm$ standard deviation).
        Top: Models trained on the fastMRI CORPD dataset.
        Bottom: Models trained on the CelebA-HQ dataset.
        Bold typeface indicates the best model per mask (\emph{G}aussian, \emph{R}adial, \emph{P}oisson) and sampling algorithm.
        Italic typeface indicates that the CelebA-HQ model outperforms the corresponding fastMRI model.
        The last three lines show the number of trainable parameters in millions, the receptive field size $r_0$ and the reconstruction time per slice.%
    }%
    \label{tab:mri}
    \resizebox{\linewidth}{!}{%
     \sisetup{table-alignment-mode = format, table-number-alignment = center, mode = text, reset-text-shape = false}\begin{tabular}{ccc*{12}{S[table-format=2.2(1),separate-uncertainty,table-align-uncertainty,detect-weight]}}\toprule
\multicolumn{15}{c}{fastMRI-trained}\\
 & & & & & \multicolumn{5}{c}{Chung \& Ye} & \multicolumn{5}{c}{Jalal~\etal}\\\mymidrule
 & & {AF} & {TV} & {U-Net} & {$d=4*$} & {$d=4$} & {$d=3$} & {$d=2$} & {$d=1$} & {$d=4*$} & {$d=4$} & {$d=3$} & {$d=2$} & {$d=1$}\\\mymidrule
\multirow{10}{*}{\rotatebox[origin=c]{90}{CORPD}}
&\multirow{2}{*}{G-1D}&\multirow{2}{*}{4}
 & 32.35 \pm 2.94 & 27.98 \pm 1.83 &32.52 \pm 2.25 & \bfseries 32.81 \pm 2.21 & 32.29 \pm 2.31 & 32.45 \pm 2.27 & 32.19 \pm 2.35
 & \bfseries 32.20 \pm 1.14 & 32.16 \pm 1.34 & 32.18 \pm 1.09 & 32.18 \pm 1.15 & 32.18 \pm 1.11\\
 & & & 0.87 \pm 0.04 & 0.76 \pm 0.05 &0.82 \pm 0.04 & \bfseries 0.83 \pm 0.04 & 0.81 \pm 0.04 & 0.82 \pm 0.04 & 0.81 \pm 0.04
 &\bfseries 0.77 \pm 0.01 & 0.77 \pm 0.02 & 0.77 \pm 0.01 & 0.77 \pm 0.01 & 0.77 \pm 0.01\\
&\multirow{2}{*}{G-1D}&\multirow{2}{*}{8}
 & 28.70 \pm 3.28 & 26.59 \pm 1.81 &31.60 \pm 2.09 & \bfseries 31.74 \pm 2.07 & 31.52 \pm 2.09 & 31.56 \pm 2.09 & 31.42 \pm 2.08
 & \bfseries 31.26 \pm 1.27 & 31.10 \pm 1.35 & 31.26 \pm 1.23 & 31.25 \pm 1.33 & 31.25 \pm 1.28\\
 & & & 0.79 \pm 0.06 & 0.71 \pm 0.06 &0.79 \pm 0.04 & \bfseries 0.79 \pm 0.04 & 0.78 \pm 0.05 & 0.78 \pm 0.04 & 0.78 \pm 0.05
 &0.73 \pm 0.02 & 0.72 \pm 0.03 & 0.73 \pm 0.02 & \bfseries 0.73 \pm 0.02 & 0.73 \pm 0.02\\
&\multirow{2}{*}{G-2D}&\multirow{2}{*}{4}
 & 32.16 \pm 2.85 & 30.26 \pm 1.73 &32.41 \pm 2.21 & \bfseries 32.57 \pm 2.17 & 32.32 \pm 2.22 & 32.42 \pm 2.20 & 32.36 \pm 2.22
 & 31.69 \pm 1.19 & 31.46 \pm 1.09 & 31.74 \pm 1.13 & 31.68 \pm 1.12 & \bfseries 31.75 \pm 1.16\\
 & & & 0.85 \pm 0.05 & 0.79 \pm 0.05 &0.80 \pm 0.05 & \bfseries 0.80 \pm 0.05 & 0.80 \pm 0.05 & 0.80 \pm 0.05 & 0.80 \pm 0.05
 &0.74 \pm 0.03 & 0.73 \pm 0.03 & 0.74 \pm 0.02 & 0.74 \pm 0.02 & \bfseries 0.74 \pm 0.02\\
&\multirow{2}{*}{R}&\multirow{2}{*}{11}
 & 29.05 \pm 2.77 & 27.92 \pm 1.85 &30.93 \pm 2.10 & \bfseries 30.99 \pm 2.10 & 30.96 \pm 2.09 & 30.93 \pm 2.09 & 30.82 \pm 2.11
 & 30.56 \pm 1.29 & 30.46 \pm 1.35 & \bfseries 30.58 \pm 1.25 & 30.50 \pm 1.32 & 30.47 \pm 1.29\\
 & & & 0.78 \pm 0.06 & 0.69 \pm 0.05 &0.75 \pm 0.05 & \bfseries 0.75 \pm 0.05 & 0.75 \pm 0.05 & 0.75 \pm 0.05 & 0.74 \pm 0.05
 &0.69 \pm 0.02 & 0.68 \pm 0.02 & \bfseries 0.69 \pm 0.02 & 0.69 \pm 0.02 & 0.69 \pm 0.02\\
&\multirow{2}{*}{P}&\multirow{2}{*}{15}
 & 25.27 \pm 2.40 & 22.74 \pm 1.46 &31.54 \pm 1.96 & \bfseries 31.58 \pm 1.97 & 31.57 \pm 1.95 & 31.48 \pm 1.97 & 31.22 \pm 1.94
 & 30.78 \pm 1.20 & 30.65 \pm 1.27 & \bfseries 30.81 \pm 1.15 & 30.72 \pm 1.24 & 30.69 \pm 1.18\\
 & & & 0.68 \pm 0.07 & 0.60 \pm 0.07 &0.76 \pm 0.05 & 0.76 \pm 0.05 & \bfseries 0.76 \pm 0.05 & 0.76 \pm 0.05 & 0.75 \pm 0.05
 &0.70 \pm 0.02 & 0.69 \pm 0.03 & \bfseries 0.70 \pm 0.02 & 0.69 \pm 0.02 & 0.69 \pm 0.02\\
\mymidrule
\multirow{10}{*}{\rotatebox[origin=c]{90}{CORPDFS}}
&\multirow{2}{*}{G-1D}&\multirow{2}{*}{4}
 & 31.47 \pm 2.69 & 30.11 \pm 1.93 &31.08 \pm 2.33 & 31.40 \pm 2.13 & 31.31 \pm 2.23 & 31.34 \pm 2.36 & \bfseries 31.70 \pm 1.98
 & 30.76 \pm 1.20 & 30.61 \pm 1.22 & 30.86 \pm 1.14 & 30.90 \pm 1.11 & \bfseries 31.07 \pm 0.99\\
 & & & 0.80 \pm 0.06 & 0.75 \pm 0.06 &0.72 \pm 0.08 & 0.74 \pm 0.07 & 0.73 \pm 0.07 & 0.73 \pm 0.08 & \bfseries 0.75 \pm 0.06
 &0.70 \pm 0.04 & 0.69 \pm 0.04 & 0.70 \pm 0.03 & 0.71 \pm 0.03 & \bfseries 0.71 \pm 0.03\\
&\multirow{2}{*}{G-1D}&\multirow{2}{*}{8}
 & 29.17 \pm 2.72 & 29.05 \pm 1.91 &29.52 \pm 2.33 & 29.76 \pm 2.23 & 29.71 \pm 2.22 & 29.70 \pm 2.35 & \bfseries 30.24 \pm 2.00
 & 29.69 \pm 1.23 & 29.47 \pm 1.26 & 29.80 \pm 1.19 & 29.87 \pm 1.15 & \bfseries 30.10 \pm 1.04\\
 & & & 0.73 \pm 0.08 & 0.70 \pm 0.07 &0.64 \pm 0.09 & 0.65 \pm 0.09 & 0.65 \pm 0.09 & 0.65 \pm 0.10 & \bfseries 0.68 \pm 0.07
 &0.64 \pm 0.04 & 0.63 \pm 0.04 & 0.64 \pm 0.04 & 0.65 \pm 0.04 & \bfseries 0.66 \pm 0.03\\
&\multirow{2}{*}{G-2D}&\multirow{2}{*}{4}
 & 30.72 \pm 2.84 & 30.73 \pm 1.89 &30.76 \pm 2.13 & 30.97 \pm 2.08 & 30.95 \pm 2.11 & 31.03 \pm 2.07 & \bfseries 31.25 \pm 1.95
 & 30.10 \pm 1.29 & 30.02 \pm 1.27 & 30.25 \pm 1.22 & 30.26 \pm 1.21 & \bfseries 30.48 \pm 1.09\\
 & & & 0.76 \pm 0.08 & 0.72 \pm 0.09 &0.68 \pm 0.09 & 0.69 \pm 0.08 & 0.69 \pm 0.08 & 0.69 \pm 0.08 & \bfseries 0.70 \pm 0.07
 &0.65 \pm 0.05 & 0.64 \pm 0.05 & 0.65 \pm 0.05 & 0.65 \pm 0.05 & \bfseries 0.66 \pm 0.04\\
&\multirow{2}{*}{R}&\multirow{2}{*}{11}
 & 29.18 \pm 2.50 & 28.75 \pm 1.71 &28.90 \pm 2.12 & 29.05 \pm 2.03 & 29.14 \pm 2.06 & 29.06 \pm 2.11 & \bfseries 29.56 \pm 1.84
 & 29.09 \pm 1.17 & 28.92 \pm 1.22 & 29.24 \pm 1.14 & 29.23 \pm 1.09 & \bfseries 29.39 \pm 1.03\\
 & & & 0.68 \pm 0.09 & 0.65 \pm 0.08 &0.58 \pm 0.10 & 0.59 \pm 0.09 & 0.60 \pm 0.09 & 0.59 \pm 0.10 & \bfseries 0.62 \pm 0.08
 &0.59 \pm 0.05 & 0.58 \pm 0.05 & 0.60 \pm 0.05 & 0.60 \pm 0.04 & \bfseries 0.60 \pm 0.04\\
&\multirow{2}{*}{P}&\multirow{2}{*}{15}
 & 28.09 \pm 2.46 & 25.51 \pm 2.00 &29.00 \pm 2.23 & 29.20 \pm 2.16 & \bfseries 29.25 \pm 2.20 & 29.17 \pm 2.20 & 29.25 \pm 2.12
 & 29.14 \pm 1.26 & 28.79 \pm 1.29 & 29.27 \pm 1.22 & 29.22 \pm 1.19 & \bfseries 29.35 \pm 1.12\\
 & & & 0.67 \pm 0.09 & 0.60 \pm 0.08 &0.59 \pm 0.10 & 0.60 \pm 0.09 & 0.61 \pm 0.09 & 0.60 \pm 0.10 & \bfseries 0.62 \pm 0.08
 &0.59 \pm 0.05 & 0.57 \pm 0.05 & 0.60 \pm 0.05 & 0.59 \pm 0.05 & \bfseries 0.60 \pm 0.04\\
\mymidrule
\multirow{10}{*}{\rotatebox[origin=c]{90}{Brain}}
&\multirow{2}{*}{G-1D}&\multirow{2}{*}{4}
 & 33.65 \pm 2.84 & 27.99 \pm 1.91 & 32.69 \pm 1.86 & 32.96 \pm 1.82 & 32.91 \pm 1.86 & \bfseries 32.96 \pm 1.79 & 32.73 \pm 1.87
 & 32.61 \pm 0.90 & 32.41 \pm 1.08 & 32.47 \pm 0.85 & 32.66 \pm 0.91 & \bfseries 32.66 \pm 0.86\\
 & & & 0.92 \pm 0.03 & 0.71 \pm 0.05 &0.87 \pm 0.03 & \bfseries 0.88 \pm 0.03 & 0.88 \pm 0.03 & 0.87 \pm 0.03 & 0.87 \pm 0.03
 &\bfseries 0.79 \pm 0.03 & 0.78 \pm 0.03 & 0.78 \pm 0.03 & 0.79 \pm 0.02 & 0.79 \pm 0.03\\
&\multirow{2}{*}{G-1D}&\multirow{2}{*}{8}
 & 30.17 \pm 3.22 & 26.30 \pm 1.90 & 30.66 \pm 2.40 & 30.98 \pm 2.37 & 31.02 \pm 2.38 & \bfseries 31.09 \pm 2.28 & 31.05 \pm 2.42
 & 31.45 \pm 1.15 & 31.25 \pm 1.27 & 31.44 \pm 1.14 & 31.58 \pm 1.13 & \bfseries 31.59 \pm 1.10\\
 & & & 0.85 \pm 0.06 & 0.68 \pm 0.06 &0.81 \pm 0.04 & 0.82 \pm 0.04 & \bfseries 0.82 \pm 0.04 & 0.82 \pm 0.04 & 0.82 \pm 0.04
 &0.75 \pm 0.03 & 0.74 \pm 0.03 & 0.75 \pm 0.03 & 0.75 \pm 0.02 & \bfseries 0.75 \pm 0.02\\
&\multirow{2}{*}{G-2D}&\multirow{2}{*}{4}
 & 31.30 \pm 2.60 & 30.27 \pm 1.65 & 32.79 \pm 1.63 & \bfseries 32.94 \pm 1.58 & 32.94 \pm 1.60 & 32.85 \pm 1.55 & 32.60 \pm 1.50
 & 32.15 \pm 0.93 & 31.86 \pm 0.94 & 32.19 \pm 1.02 & \bfseries 32.25 \pm 0.91 & 32.20 \pm 0.82\\
 & & & 0.91 \pm 0.04 & 0.79 \pm 0.04 &0.86 \pm 0.03 & 0.87 \pm 0.03 & \bfseries 0.87 \pm 0.03 & 0.87 \pm 0.03 & 0.86 \pm 0.03
 &0.77 \pm 0.04 & 0.76 \pm 0.04 & 0.77 \pm 0.04 & \bfseries 0.77 \pm 0.03 & 0.77 \pm 0.03\\
&\multirow{2}{*}{R}&\multirow{2}{*}{11}
 & 28.86 \pm 2.83 & 26.01 \pm 1.70 & 30.14 \pm 2.24 & 30.30 \pm 2.24 & 30.43 \pm 2.19 & 30.27 \pm 2.12 & \bfseries 30.43 \pm 2.17
 & 30.84 \pm 1.09 & 30.69 \pm 1.18 & \bfseries 30.91 \pm 1.04 & 30.89 \pm 1.10 & 30.86 \pm 1.07\\
 & & & 0.83 \pm 0.06 & 0.60 \pm 0.05 &0.78 \pm 0.05 & 0.79 \pm 0.05 & \bfseries 0.80 \pm 0.04 & 0.79 \pm 0.04 & 0.79 \pm 0.05
 &0.72 \pm 0.03 & 0.72 \pm 0.03 & 0.73 \pm 0.03 & \bfseries 0.73 \pm 0.03 & 0.73 \pm 0.03\\
&\multirow{2}{*}{P}&\multirow{2}{*}{15}
 & 27.92 \pm 3.28 & 23.33 \pm 1.62 & 30.71 \pm 2.55 & 30.72 \pm 2.64 & \bfseries 31.02 \pm 2.50 & 30.88 \pm 2.42 & 30.95 \pm 2.55
 & 31.25 \pm 1.06 & 31.01 \pm 1.26 & \bfseries 31.29 \pm 1.00 & 31.28 \pm 1.06 & 31.25 \pm 1.04\\
 & & & 0.72 \pm 0.08 & 0.52 \pm 0.05 &0.78 \pm 0.06 & 0.77 \pm 0.06 & \bfseries 0.79 \pm 0.05 & 0.78 \pm 0.05 & 0.79 \pm 0.06
 &0.73 \pm 0.03 & 0.72 \pm 0.03 & 0.73 \pm 0.03 & \bfseries 0.73 \pm 0.03 & 0.73 \pm 0.03\\
\midrule
\multicolumn{15}{c}{CelebA-trained}\\
 & & & & & \multicolumn{5}{c}{Chung \& Ye} & \multicolumn{5}{c}{Jalal~\etal}\\\mymidrule
 & & {AF} & {TV} & {U-Net} & {$d=4*$} & {$d=4$} & {$d=3$} & {$d=2$} & {$d=1$} & {$d=4*$} & {$d=4$} & {$d=3$} & {$d=2$} & {$d=1$}\\\mymidrule
\multirow{10}{*}{\rotatebox[origin=c]{90}{CORPD}}
&\multirow{2}{*}{G-1D}&\multirow{2}{*}{4}
 & {\textemdash} & {\textemdash} & 32.02 \pm 2.26 & 32.35 \pm 2.29 & 31.57 \pm 2.41 & 32.29 \pm 2.37 & \bfseries \itshape 32.61 \pm 2.19
 &32.14 \pm 1.19 & 32.09 \pm 1.21 & 32.00 \pm 1.23 & \itshape 32.18 \pm 1.24 & \bfseries \itshape 32.27 \pm 1.32\\
 & & & {\textemdash} & {\textemdash} &\itshape 0.82 \pm 0.03 & \itshape 0.83 \pm 0.03 & \itshape 0.82 \pm 0.04 & \itshape 0.83 \pm 0.03 & \bfseries \itshape 0.84 \pm 0.03
 &0.77 \pm 0.03 & 0.77 \pm 0.03 & 0.77 \pm 0.03 & \itshape 0.77 \pm 0.03 & \bfseries \itshape 0.78 \pm 0.03\\
&\multirow{2}{*}{G-1D}&\multirow{2}{*}{8}
 & {\textemdash} & {\textemdash} & 31.58 \pm 1.92 & 31.63 \pm 1.90 & \bfseries \itshape 31.78 \pm 1.93 & \itshape 31.67 \pm 1.86 & \itshape 31.58 \pm 1.88
 &31.03 \pm 1.16 & 31.09 \pm 1.50 & \bfseries 31.17 \pm 1.51 & 31.13 \pm 1.43 & 31.13 \pm 1.32\\
 & & & {\textemdash} & {\textemdash} &\itshape 0.79 \pm 0.04 & \itshape 0.79 \pm 0.04 & \bfseries \itshape 0.80 \pm 0.04 & \itshape 0.80 \pm 0.04 & \itshape 0.80 \pm 0.03
 &0.72 \pm 0.03 & \itshape 0.72 \pm 0.03 & \bfseries \itshape 0.73 \pm 0.03 & 0.73 \pm 0.03 & 0.73 \pm 0.03\\
&\multirow{2}{*}{G-2D}&\multirow{2}{*}{4}
 & {\textemdash} & {\textemdash} & \itshape 33.70 \pm 1.88 & \bfseries \itshape 33.73 \pm 1.86 & \itshape 33.63 \pm 1.94 & \itshape 33.57 \pm 1.94 & \itshape 33.54 \pm 1.93
 &31.31 \pm 0.75 & \itshape 31.54 \pm 0.98 & 31.34 \pm 0.77 & \bfseries 31.56 \pm 0.89 & 31.50 \pm 0.93\\
 & & & {\textemdash} & {\textemdash} &\itshape 0.84 \pm 0.03 & \bfseries \itshape 0.84 \pm 0.03 & \itshape 0.84 \pm 0.03 & \itshape 0.84 \pm 0.03 & \itshape 0.84 \pm 0.03
 &0.72 \pm 0.01 & \itshape 0.73 \pm 0.01 & 0.73 \pm 0.01 & \bfseries 0.74 \pm 0.01 & 0.73 \pm 0.01\\
&\multirow{2}{*}{R}&\multirow{2}{*}{11}
 & {\textemdash} & {\textemdash} & 30.78 \pm 2.07 & 30.78 \pm 2.10 & \bfseries 30.96 \pm 2.03 & 30.78 \pm 2.07 & 30.68 \pm 2.05
 &30.35 \pm 1.22 & 30.31 \pm 1.21 & 30.29 \pm 1.21 & \bfseries 30.37 \pm 1.25 & 30.27 \pm 1.23\\
 & & & {\textemdash} & {\textemdash} &0.74 \pm 0.05 & 0.74 \pm 0.05 & \bfseries \itshape 0.75 \pm 0.05 & 0.75 \pm 0.05 & 0.74 \pm 0.05
 &0.68 \pm 0.02 & 0.67 \pm 0.02 & 0.67 \pm 0.02 & \bfseries 0.68 \pm 0.02 & 0.67 \pm 0.02\\
&\multirow{2}{*}{P}&\multirow{2}{*}{15}
 & {\textemdash} & {\textemdash} & \bfseries 30.37 \pm 2.02 & 30.35 \pm 2.03 & 30.29 \pm 2.06 & 29.71 \pm 2.22 & 28.62 \pm 2.60
 &30.38 \pm 1.08 & \bfseries 30.42 \pm 1.22 & 30.20 \pm 1.25 & 30.42 \pm 1.36 & 30.34 \pm 1.21\\
 & & & {\textemdash} & {\textemdash} &0.73 \pm 0.05 & 0.73 \pm 0.05 & \bfseries 0.73 \pm 0.05 & 0.72 \pm 0.05 & 0.70 \pm 0.06
 &0.68 \pm 0.02 & 0.68 \pm 0.03 & 0.67 \pm 0.03 & \bfseries 0.69 \pm 0.03 & 0.68 \pm 0.02\\
\mymidrule
\multirow{10}{*}{\rotatebox[origin=c]{90}{CORPDFS}}
&\multirow{2}{*}{G-1D}&\multirow{2}{*}{4}
 & {\textemdash} & {\textemdash} &\itshape 31.63 \pm 1.95 & 31.29 \pm 1.98 & 31.17 \pm 2.00 & \itshape 31.78 \pm 2.06 & \bfseries \itshape 31.86 \pm 1.99
 &\itshape 31.35 \pm 0.99 & \bfseries \itshape 31.40 \pm 1.08 & \itshape 31.28 \pm 0.93 & \itshape 31.30 \pm 0.93 & \itshape 31.28 \pm 1.02\\
 & & & {\textemdash} & {\textemdash} &\itshape 0.76 \pm 0.05 & \itshape 0.76 \pm 0.06 & \itshape 0.76 \pm 0.06 & \itshape 0.77 \pm 0.06 & \bfseries \itshape 0.77 \pm 0.06
 &\itshape 0.72 \pm 0.03 & \bfseries \itshape 0.73 \pm 0.03 & \itshape 0.72 \pm 0.03 & \itshape 0.72 \pm 0.03 & \itshape 0.72 \pm 0.03\\
&\multirow{2}{*}{G-1D}&\multirow{2}{*}{8}
 & {\textemdash} & {\textemdash} &\itshape 31.04 \pm 2.00 & \itshape 31.02 \pm 2.01 & \itshape 31.10 \pm 2.01 & \bfseries \itshape 31.10 \pm 2.02 & \itshape 31.00 \pm 1.99
 &\itshape 30.38 \pm 0.85 & \itshape 30.38 \pm 0.98 & \itshape 30.28 \pm 1.05 & \bfseries \itshape 30.46 \pm 1.09 & \itshape 30.41 \pm 1.04\\
 & & & {\textemdash} & {\textemdash} &\itshape 0.71 \pm 0.07 & \itshape 0.71 \pm 0.07 & \bfseries \itshape 0.72 \pm 0.07 & \itshape 0.72 \pm 0.07 & \itshape 0.71 \pm 0.07
 &\itshape 0.67 \pm 0.03 & \itshape 0.67 \pm 0.04 & \itshape 0.67 \pm 0.04 & \bfseries \itshape 0.68 \pm 0.04 & \itshape 0.67 \pm 0.04\\
&\multirow{2}{*}{G-2D}&\multirow{2}{*}{4}
 & {\textemdash} & {\textemdash} &\bfseries \itshape 31.98 \pm 2.12 & \itshape 31.76 \pm 2.09 & \itshape 31.83 \pm 2.08 & \itshape 31.84 \pm 2.09 & \itshape 31.78 \pm 2.07
 &\itshape 30.48 \pm 0.84 & \itshape 30.45 \pm 0.97 & \itshape 30.44 \pm 0.83 & \bfseries \itshape 30.58 \pm 0.95 & \itshape 30.50 \pm 0.97\\
 & & & {\textemdash} & {\textemdash} &\bfseries \itshape 0.73 \pm 0.08 & \itshape 0.73 \pm 0.08 & \itshape 0.73 \pm 0.08 & \itshape 0.73 \pm 0.08 & \itshape 0.73 \pm 0.08
 &\itshape 0.67 \pm 0.03 & \itshape 0.66 \pm 0.04 & \itshape 0.66 \pm 0.03 & \bfseries \itshape 0.67 \pm 0.04 & \itshape 0.67 \pm 0.04\\
&\multirow{2}{*}{R}&\multirow{2}{*}{11}
 & {\textemdash} & {\textemdash} &\itshape 30.30 \pm 1.92 & \itshape 30.29 \pm 1.93 & \bfseries \itshape 30.36 \pm 1.93 & \itshape 30.30 \pm 1.92 & \itshape 30.20 \pm 1.91
 &\bfseries \itshape 29.72 \pm 0.93 & \itshape 29.69 \pm 0.92 & \itshape 29.60 \pm 0.96 & \itshape 29.72 \pm 0.96 & \itshape 29.63 \pm 0.91\\
 & & & {\textemdash} & {\textemdash} &\itshape 0.66 \pm 0.08 & \itshape 0.66 \pm 0.08 & \bfseries \itshape 0.66 \pm 0.08 & \itshape 0.66 \pm 0.08 & \itshape 0.65 \pm 0.08
 &\itshape 0.62 \pm 0.04 & \itshape 0.61 \pm 0.03 & \itshape 0.61 \pm 0.04 & \bfseries \itshape 0.62 \pm 0.04 & \itshape 0.61 \pm 0.03\\
&\multirow{2}{*}{P}&\multirow{2}{*}{15}
 & {\textemdash} & {\textemdash} &\bfseries \itshape 30.08 \pm 2.06 & \itshape 30.02 \pm 2.07 & \itshape 30.03 \pm 2.03 & \itshape 29.71 \pm 2.10 & 29.19 \pm 2.18
 &\itshape 29.78 \pm 0.91 & \itshape 29.63 \pm 0.98 & \itshape 29.49 \pm 0.92 & \bfseries \itshape 29.83 \pm 1.04 & \itshape 29.69 \pm 0.89\\
 & & & {\textemdash} & {\textemdash} &\itshape 0.66 \pm 0.08 & \itshape 0.66 \pm 0.08 & \bfseries \itshape 0.66 \pm 0.08 & \itshape 0.65 \pm 0.08 & \itshape 0.64 \pm 0.08
 &\itshape 0.62 \pm 0.04 & \itshape 0.61 \pm 0.04 & \itshape 0.61 \pm 0.04 & \bfseries \itshape 0.62 \pm 0.05 & \itshape 0.62 \pm 0.04\\
\mymidrule
\multirow{10}{*}{\rotatebox[origin=c]{90}{Brain}}
&\multirow{2}{*}{G-1D}&\multirow{2}{*}{4}
 & {\textemdash} & {\textemdash} & 32.10 \pm 1.62 & 31.41 \pm 1.89 & 31.26 \pm 1.51 & 32.32 \pm 1.78 & \bfseries 32.61 \pm 1.52
 &\itshape 32.73 \pm 1.26 & \itshape 32.81 \pm 1.30 & \itshape 32.70 \pm 1.26 & \itshape 32.74 \pm 1.35 & \bfseries \itshape 32.97 \pm 1.48\\
 & & & {\textemdash} & {\textemdash} &\itshape 0.87 \pm 0.02 & 0.87 \pm 0.03 & 0.87 \pm 0.02 & \itshape 0.88 \pm 0.02 & \bfseries \itshape 0.89 \pm 0.02
 &0.78 \pm 0.06 & \itshape 0.79 \pm 0.06 & \itshape 0.78 \pm 0.06 & 0.79 \pm 0.05 & \bfseries \itshape 0.80 \pm 0.05\\
&\multirow{2}{*}{G-1D}&\multirow{2}{*}{8}
 & {\textemdash} & {\textemdash} & \itshape 31.68 \pm 2.25 & \itshape 31.45 \pm 2.40 & \itshape 31.56 \pm 1.98 & \bfseries \itshape 31.84 \pm 2.23 & \itshape 31.45 \pm 2.33
 &\bfseries \itshape 31.90 \pm 1.45 & \itshape 31.72 \pm 1.38 & \itshape 31.62 \pm 1.34 & \itshape 31.72 \pm 1.18 & \itshape 31.62 \pm 1.28\\
 & & & {\textemdash} & {\textemdash} &\itshape 0.84 \pm 0.04 & \itshape 0.84 \pm 0.04 & \bfseries \itshape 0.85 \pm 0.03 & \itshape 0.85 \pm 0.03 & \itshape 0.84 \pm 0.04
 &\bfseries \itshape 0.76 \pm 0.05 & \itshape 0.75 \pm 0.05 & \itshape 0.75 \pm 0.05 & \itshape 0.75 \pm 0.05 & 0.75 \pm 0.06\\
&\multirow{2}{*}{G-2D}&\multirow{2}{*}{4}
 & {\textemdash} & {\textemdash} & \bfseries \itshape 34.05 \pm 1.50 & \itshape 33.68 \pm 1.57 & \itshape 33.60 \pm 1.40 & \itshape 33.32 \pm 1.55 & \itshape 33.30 \pm 1.38
 &31.43 \pm 0.50 & 31.81 \pm 0.54 & 31.61 \pm 0.47 & 31.83 \pm 0.53 & \bfseries 31.89 \pm 0.57\\
 & & & {\textemdash} & {\textemdash} &\bfseries \itshape 0.89 \pm 0.02 & \itshape 0.89 \pm 0.02 & \itshape 0.89 \pm 0.02 & \itshape 0.89 \pm 0.02 & \itshape 0.89 \pm 0.02
 &0.73 \pm 0.04 & 0.75 \pm 0.04 & 0.74 \pm 0.04 & 0.75 \pm 0.04 & \bfseries 0.75 \pm 0.04\\
&\multirow{2}{*}{R}&\multirow{2}{*}{11}
 & {\textemdash} & {\textemdash} & \itshape 31.29 \pm 2.34 & \itshape 31.19 \pm 2.35 & \bfseries \itshape 31.39 \pm 2.28 & \itshape 31.24 \pm 2.30 & \itshape 31.01 \pm 2.36
 &\itshape 31.06 \pm 0.93 & \itshape 31.06 \pm 0.93 & \itshape 30.98 \pm 0.97 & \bfseries \itshape 31.09 \pm 0.95 & \itshape 31.00 \pm 0.96\\
 & & & {\textemdash} & {\textemdash} &\itshape 0.81 \pm 0.05 & \itshape 0.81 \pm 0.05 & \bfseries \itshape 0.81 \pm 0.04 & \itshape 0.81 \pm 0.04 & \itshape 0.80 \pm 0.05
 &0.72 \pm 0.03 & \itshape 0.72 \pm 0.03 & 0.72 \pm 0.03 & \bfseries 0.72 \pm 0.03 & 0.72 \pm 0.03\\
&\multirow{2}{*}{P}&\multirow{2}{*}{15}
 & {\textemdash} & {\textemdash} & \itshape 31.71 \pm 2.56 & \itshape 31.70 \pm 2.55 & \bfseries \itshape 31.78 \pm 2.55 & \itshape 31.16 \pm 2.97 & 30.64 \pm 3.27
 &\bfseries \itshape 31.35 \pm 0.89 & \itshape 31.31 \pm 0.91 & 31.27 \pm 1.30 & \itshape 31.32 \pm 1.04 & \itshape 31.29 \pm 0.97\\
 & & & {\textemdash} & {\textemdash} &\itshape 0.80 \pm 0.05 & \itshape 0.80 \pm 0.05 & \bfseries \itshape 0.81 \pm 0.05 & 0.78 \pm 0.07 & 0.76 \pm 0.09
 &\bfseries 0.73 \pm 0.03 & \itshape 0.73 \pm 0.03 & 0.73 \pm 0.05 & 0.73 \pm 0.04 & 0.73 \pm 0.03\\
 \midrule 
 \multicolumn{3}{c}{\# params (M)} & {0} & {496} & {61.4} & {61.2} & {42.4} & {23.9} & {5.3} & {61.4} & {61.2} & {42.4} & {23.9} & {5.3}\\
 \multicolumn{3}{c}{\( r_0 \)} & {2} & {>640} & {>640} & {>640} & {331} & {143} & {49} & {>640} & {>640} & {331} & {143} & {49}\\
  \multicolumn{3}{c}{time in \unit{\second}} & {0.42} & \num{5e-3}  & {84.4} & {84.18} & {81.51} & {74.28} & {39.23} & {22.17} & {22.06} & {21.35} & {19.46} & {10.29}\\
\bottomrule
\end{tabular}
        } %
\end{table}

\robustify{\bfseries}
\robustify{\itshape}

\begin{table}
    \def\mymidrule{\cmidrule(r{.5em}){1-3}\cmidrule(l{.5em}){4-8}}
    \centering
    \caption{%
        Quantitative reconstruction results in terms of PSNR and SSIM (mean $\pm$ standard deviation).
        Top: Models trained on the thoracic CT dataset.
        Bottom: Models trained on the CelebA-HQ dataset.
        Bold typeface indicates the best model per mask (\emph{G}aussian, \emph{R}adial, \emph{P}oisson) and sampling algorithm.
        Italic typeface indicates that the CelebA-HQ model outperforms the corresponding CT model.
        The last three lines show the number of trainable parameters in millions, the receptive field size $r_0$ and the reconstruction time per slice.%
    }%
    \label{tab:ct}
    \resizebox{\linewidth}{!}{%
     \sisetup{table-alignment-mode = format, table-number-alignment = center, mode = text, reset-text-shape = false}\begin{tabular}{cc*{6}{S[table-format=2.2(1),separate-uncertainty,table-align-uncertainty,detect-weight]}}\toprule\multicolumn{8}{c}{CT-trained}\\
 & & & \multicolumn{5}{c}{Chung \& Ye} \\\mymidrule
& {$N_\theta$} & {TV} & {$d=4*$} & {$d=4$} & {$d=3$} & {$d=2$} & {$d=1$}\\\mymidrule
\footnotesize \multirow{4}{*}{\scriptsize\rotatebox[origin=c]{90}{Thorax CT}}
&\multirow{2}{*}{60}
 & 23.21 \pm 5.55 &36.29 \pm 1.65 & \bfseries 36.30 \pm 1.64 & 36.29 \pm 1.64 & 36.27 \pm 1.63 & 36.14 \pm 1.63\\
 & & 0.54 \pm 0.25 &0.85 \pm 0.05 & \bfseries 0.85 \pm 0.05 & 0.85 \pm 0.05 & 0.85 \pm 0.05 & 0.85 \pm 0.05\\
&\multirow{2}{*}{30}
 & 28.45 \pm 3.76 &35.16 \pm 1.70 & \bfseries 35.22 \pm 1.70 & 35.18 \pm 1.69 & 35.13 \pm 1.69 & 34.92 \pm 1.69\\
 & & 0.83 \pm 0.09 &0.83 \pm 0.05 & \bfseries 0.83 \pm 0.05 & 0.83 \pm 0.05 & 0.83 \pm 0.05 & 0.83 \pm 0.05\\
\mymidrule
\multirow{4}{*}{\scriptsize\rotatebox[origin=c]{90}{Head CT}}
&\multirow{2}{*}{60}
 & 29.58 \pm 2.45 &34.85 \pm 1.29 & 34.87 \pm 1.30 & \bfseries 34.93 \pm 1.30 & 34.78 \pm 1.29 & 34.76 \pm 1.28\\
 & & 0.47 \pm 0.17 &0.89 \pm 0.01 & 0.89 \pm 0.01 & \bfseries 0.89 \pm 0.01 & 0.89 \pm 0.01 & 0.88 \pm 0.01\\
&\multirow{2}{*}{30}
 & 31.21 \pm 2.71 &31.58 \pm 1.82 & 31.68 \pm 1.85 & \bfseries 31.79 \pm 1.85 & 31.40 \pm 1.88 & 31.29 \pm 1.83\\
 & & 0.94 \pm 0.02 &0.79 \pm 0.05 & 0.80 \pm 0.05 & \bfseries 0.81 \pm 0.05 & 0.78 \pm 0.06 & 0.78 \pm 0.06\\
 \midrule
 \multicolumn{8}{c}{CelebA-trained}\\
 & & & \multicolumn{5}{c}{Chung \& Ye} \\\mymidrule
& {$N_\theta$} & {TV} & {$d=4*$} & {$d=4$} & {$d=3$} & {$d=2$} & {$d=1$}\\\mymidrule
\multirow{4}{*}{\scriptsize\rotatebox[origin=c]{90}{Thorax CT}}
&\multirow{2}{*}{60}
 & {---} &\bfseries 35.70 \pm 1.63 & 35.70 \pm 1.63 & 35.69 \pm 1.62 & 35.69 \pm 1.63 & 35.54 \pm 1.62\\
 & & {---} &0.84 \pm 0.05 & \bfseries 0.84 \pm 0.05 & 0.84 \pm 0.05 & 0.84 \pm 0.05 & 0.84 \pm 0.05\\
&\multirow{2}{*}{30}
 & {---} &34.17 \pm 1.75 & 34.21 \pm 1.75 & \bfseries 34.25 \pm 1.74 & 34.20 \pm 1.75 & 33.99 \pm 1.76\\
 & & {---} &0.81 \pm 0.05 & 0.81 \pm 0.05 & \bfseries 0.81 \pm 0.05 & 0.81 \pm 0.05 & 0.80 \pm 0.05\\
\mymidrule
\multirow{4}{*}{\scriptsize\rotatebox[origin=c]{90}{Head CT}}
&\multirow{2}{*}{60}
 & {---} & \itshape 35.13 \pm 1.27 & \itshape 35.11 \pm 1.24 & \itshape \bfseries 35.14 \pm 1.28 & \itshape 35.14 \pm 1.27 & \itshape 34.95 \pm 1.25\\
 & & {---} & \itshape 0.89 \pm 0.01 & \itshape 0.89 \pm 0.01 & \itshape \bfseries 0.89 \pm 0.01 & \itshape 0.89 \pm 0.01 & \itshape 0.89 \pm 0.01\\
&\multirow{2}{*}{30}
 & {---} & \itshape 31.96 \pm 1.93 & \itshape 31.85 \pm 1.92 & \itshape \bfseries 32.17 \pm 1.89 & \itshape 32.00 \pm 1.90 & \itshape 31.63 \pm 1.91\\
 & & {---} &\itshape 0.81 \pm 0.05 & \itshape 0.81 \pm 0.05 & \itshape \bfseries 0.82 \pm 0.04 & \itshape 0.82 \pm 0.04 & \itshape 0.79 \pm 0.06\\
 \midrule 
 \multicolumn{2}{c}{\# params (M)} & {0} & {61.4} & {61.2} & {42.4} & {23.9} & {5.3}\\
 \multicolumn{2}{c}{\( r_0 \)} & {2} & {>640} & {>640} & {331} & {143} & {49}\\
  \multicolumn{2}{c}{time in \unit{\second}} & {37.85} & {207.8} & {208.6} & {204} & {189} & {118.4}\\
 \bottomrule
\end{tabular}
        } %
\end{table}

\end{document}